\documentclass[a4paper,11pt]{article}
\pdfoutput=1 
\usepackage{jcappub} 
\usepackage[T1]{fontenc} 
\usepackage{longtable}
\usepackage{graphicx}

\title{\boldmath 
Enhancement of primordial curvature perturbations in $R^3$-corrected Starobinsky-Higgs inflation
}

\author[a]{Jinsu Kim,}
\author[a,b,c]{Xinpeng Wang,}
\author[a,b,d,e,f]{Ying-li Zhang,}
\author[a]{and Zhongzhou Ren}

\affiliation[a]{
	School of Physics Science and Engineering,
	Tongji University,\\
	Shanghai 200092, China
}
\affiliation[b]{
    Kavli Institute for the Physics and Mathematics of the Universe (WPI),\\
    The University of Tokyo Institutes for Advanced Study,
    The University of Tokyo, \\
    Chiba 277-8583, Japan
}
\affiliation[c]{
  Department of Physics, 
  Graduate School of Science, 
  The University of Tokyo, \\
  Tokyo 113-0033, Japan
}

\affiliation[d]{
    Institute for Advanced Study of Tongji University, \\
    Shanghai 200092, China
}
\affiliation[e]{
    Institute of Theoretical Physics,
    Chinese Academy of Sciences, \\
    Beijing 100190, China
}

\affiliation[f]{
    Center for Gravitation and Cosmology,
    Yangzhou University, \\
    Yangzhou 225009, China
}

\emailAdd{kimjinsu@tongji.edu.cn}
\emailAdd{xinpengwang@tongji.edu.cn}
\emailAdd{yingli@tongji.edu.cn}

\abstract{
    We provide a systematic study of the Starobinsky-Higgs inflation model in the presence of an additional cubic term of the Ricci scalar. We investigate, in particular, the effects of the cubic term on the spectral index $n_s$ and the tensor-to-scalar ratio $r$. Through both analytical and numerical analyses, we show that the $R^3$-corrected Starobinsky-Higgs model can achieve compatibility with cosmic microwave background observations while producing distinct observational signatures with different frequency ranges. In addition, we discuss the complementarity between different observational probes, including the scalar-induced gravitational waves and spectral distortions, offering an independent probe of the enhanced curvature perturbations. Detection prospects are also discussed.
}

\begin{document}
\maketitle
\flushbottom

\section{Introduction}
\label{sec:intro}
The accelerating expansion phase of our early Universe, dubbed cosmic inflation, has become the leading paradigm for the theory of the early Universe~\cite{Brout:1977ix,Starobinsky:1980te,Guth:1980zm,Linde:1981mu,Albrecht:1982wi}. Inflation not only solves the problems of the standard Hot Big Bang cosmology, but also sets the ground for the late-time large-scale structures through tiny quantum fluctuations. One of the most important predictions of cosmic inflation is the nearly scale-invariant curvature power spectrum which is in excellent agreement with the Cosmic Microwave Background (CMB) observation such as the Planck observation~\cite{Planck:2018jri}. The simple power-law power spectrum of the form $\mathcal{P}_\mathcal{R} = A_s(k/k_*)^{n_s-1}$ is usually adopted to describe the inflationary curvature perturbation, where the tilt of the spectrum, or the spectral index, $n_s$, that characterises the deviation from the perfect scale invariance, is constrained to be $0.958 \leq n_s \leq 0.975$ (95\% C.L.)~\cite{Planck:2018jri,BICEP:2021xfz}.\footnote{
The Atacama Cosmology Telescope (ACT) collaboration has recently reported that when combined with the Planck data as well as the baryon acoustic oscillation data, a slightly larger value of the spectral index, $n_s = 0.974 \pm 0.003$, is preferred \cite{ACT:2025fju,ACT:2025tim}.
} Various inflation models have been suggested and thoroughly studied; see, for example, Ref.~\cite{Martin:2013tda}.

Another critical inflationary observable is the ratio between the power spectrum of primordial tensor perturbation $\mathcal{P}_{\rm T}$ and that of the scalar perturbation $\mathcal{P}_{\mathcal{R}}$, the so-called tensor-to-scalar ratio $r\equiv\mathcal{P}_{\rm T}/\mathcal{P}_{\mathcal{R}}$. A stringent upper bound on $r$, namely $r \leq 0.036$ (95\% C.L.), is put by the latest analysis of the Planck and BICEP/Keck data~\cite{Planck:2018jri,BICEP:2021xfz}. Such a strong bound on $r$ plays a crucial role in ruling out numerous single-field inflationary models. In particular, the chaotic inflationary model with the inflaton potential of the form $V(\phi) \sim \phi^n$ is ruled out. However, there exist mechanisms to bring the chaotic inflationary model to the observationally-favoured region. One of such mechanisms is to introduce a coupling of the inflaton field to gravity~\cite{Futamase:1987ua,Cervantes-Cota:1995ehs,Bezrukov:2007ep}. The so-called non-minimal coupling of the inflaton field to the Ricci scalar of the form $\xi \phi^2 R$ is known to have the effect of flattening the scalar potential, reducing the tensor-to-scalar ratio. The Higgs inflationary model is one good example that adopts such a non-minimal coupling~\cite{Bezrukov:2007ep,Bezrukov:2010jz}; see Refs.~\cite{Rubio:2018ogq,Cheong:2021vdb} for a review and Refs.~\cite{Einhorn:2009bh,Ferrara:2010yw,Ferrara:2010in,Pallis:2011gr,Arai:2011nq,Einhorn:2012ih,Arai:2011aa,Arai:2012em,Kawai:2014doa,Kawai:2014gqa,Kawai:2015ryj,Kawai:2022emp} for supersymmetric extensions of Higgs inflation. See also Refs.~\cite{Ferrara:2013rsa,Kallosh:2013yoa,Hyun:2022uzc,Hyun:2023bkf} for other mechanisms.
The prediction of Higgs inflation on the spectral index and the tensor-to-scalar ratio is in fact very similar to that of the $R^2$ model, also known as the Starobinsky model~\cite{Nariai:1971sv,Starobinsky:1980te}. The non-minimal coupling model and the $R^2$ model are considered the most favoured inflationary models from the Planck observational data~\cite{Planck:2018jri}, although the recent ACT result disfavours these two models \cite{ACT:2025fju,ACT:2025tim}.
Both the non-minimal coupling term $\xi \phi^2 R$ and the Ricci scalar-squared term $R^2$ are of the mass dimension four. It is thus natural to include both terms on an equal footing. This scenario, commonly referred to as the Higgs-$R^2$, Starobinsky-Higgs, or Higgs-scalaron model, has been extensively studied; see, {\it e.g.}, Refs.~\cite{Salvio:2015kka,Calmet:2016fsr,Ema:2017rqn,Wang:2017fuy,He:2018gyf,Ghilencea:2018rqg,Gorbunov:2018llf,Gundhi:2018wyz,Bezrukov:2019ylq,Kim:2025ikw}.

Recently, the Higgs-$R^2$ model has attracted much attention, especially in the context of primordial black hole (PBH) formations~\cite{Pi:2017gih,Cheong:2019vzl,Gundhi:2020zvb,Cheong:2022gfc,Wang:2024vfv,Kohri:2025lau}. In particular, it has been shown that the model may realise a two-step inflationary scenario~\cite{Pi:2017gih,Cheong:2022gfc,Wang:2024vfv}; see also Refs.~\cite{Braglia:2020eai,Braglia:2020taf}. In this scenario, the inflationary trajectory first follows the scalaron direction, realising the first phase of inflation. Once the scalaron hits the local minimum, the accelerating expansion briefly halts, and the effective mass-squared of the Higgs field momentarily turns to be tachyonic. As a result, the Higgs field starts to slowly roll down its potential, realising the second phase of inflation. The two-step inflationary phase then ends when both the fields come to their global minimum. Interestingly, it has been observed that the curvature power spectrum may get enhanced during the intermediate break between the two phases of inflation as a consequence of the slow-roll violation as well as the isocurvature-to-curvature conversion~\cite{Pi:2017gih,Cheong:2022gfc,Wang:2024vfv}.
When the enhancement of the curvature power spectrum is large enough, PBHs may copiously be formed due to the gravitational collapse when the corresponding modes re-enter the horizon~\cite{Zeldovich:1967lct,Hawking:1971ei,Carr:1974nx,Polnarev:1985btg}; see, {\it e.g.}, Refs.~\cite{Carr:2020xqk,Khlopov:2008qy,Sasaki:2018dmp,Carr:2020gox,Green:2020jor,Villanueva-Domingo:2021spv,Escriva:2022duf} for a comprehensive review on PBHs.
One consequence of the enhanced curvature perturbations is that they may source the gravitational waves (GWs) at the nonlinear order, producing the so-called scalar-induced, second-order GWs~\cite{Matarrese:1997ay,Mollerach:2003nq,Ananda:2006af,Baumann:2007zm,Saito:2008jc}; see, {\it e.g.}, Refs.~\cite{Sasaki:2018dmp,Gong:2019mui,Yuan:2021qgz,Domenech:2021ztg,Domenech:2019quo,Domenech:2020kqm} for a review. The produced induced GWs may account for the recent observation of stochastic GW backgrounds by North American Nanohertz Observatory for Gravitational Waves (NANOGrav)~\cite{NANOGrav:2023gor,NANOGrav:2023hde} or fall into the frequency region to be probed by future GW experiments such as Laser Interferometer Space Antenna (LISA)~\cite{LISA:2017pwj,Baker:2019nia,LISACosmologyWorkingGroup:2024hsc,LISACosmologyWorkingGroup:2025vdz}, Deci-hertz Interferometer Gravitational wave Observatory (DECIGO)~\cite{Seto:2001qf,Kawamura:2006up,Sato:2017dkf,Isoyama:2018rjb,Kawamura:2020pcg}, Big Bang Observer (BBO)~\cite{Corbin:2005ny,Crowder:2005nr,Harry:2006fi}, TianQin~\cite{TianQin:2015yph}, and Taiji~\cite{Ruan:2018tsw}.

Although the two-step inflationary scenario in the Higgs-$R^2$ model features plenty of amusing phenomena such as the formation of PBHs and the induced GW production, it predicts the spectral index $n_s$ that is outside the bound of the latest Planck and ACT observations if the produced PBHs are to account for the whole dark matter abundance today~ \cite{Pi:2017gih,Wang:2024vfv}. The authors of Refs.~\cite{Pi:2017gih,Wang:2024vfv} suggested as a remedy the inclusion of a small, but non-negligible, cubic term of the Ricci scalar, $R^3$. They showed that a negative Ricci cubic term may shift the spectral index towards the observationally-favoured region. Based on this observation, in this work, we aim to provide a systematic study of the Higgs-$R^2$ inflationary model in the presence of the $R^3$ term. In particular, we investigate the effects of the cubic term on the spectral index $n_s$ and the tensor-to-scalar ratio $r$. Furthermore, we discuss the consequences of the inclusion of the $R^3$ term on the formation of PBHs and scalar-induced GWs. Finally, we comment on the forecast of spectral distortions.

The paper is organised as follows.
In Sec.~\ref{sec:model}, we introduce the model under consideration, setting up the conventions and notations. Closely following Ref.~\cite{Wang:2024vfv}, the inflationary dynamics is first analytically studied, for completeness. We discuss in detail how the presence of the cubic term of the Ricci scalar may affect the prediction of the Starobinsky-Higgs inflationary model, paying extra attention to the spectral index $n_s$ and the tensor-to-scalar ratio $r$. We then provide numerical treatments of the system, verifying our analytical understanding.
In Sec.~\ref{sec:PBHandGW}, we study the formation of PBHs and scalar-induced GWs, highlighting the effect of the $R^3$ term.
We comment on the forecast of spectral distortions in Sec.~\ref{sec:SD}.
We conclude in Sec.~\ref{sec:conc}.

\section{Inflationary Dynamics}
\label{sec:model}
The concrete model we consider in this work is given by
\begin{align}
    S = \int d^4x \, \sqrt{-g} \, \left\{
    \frac{M_{\rm P}^2}{2}\left[
    1 - \xi\frac{(\chi-\chi_0)^2}{M_{\rm P}^2}
    + \frac{R}{6M^2}
    + q \frac{R^2}{3M^4}
    \right]R
    -\frac{1}{2}g^{\mu\nu}\partial_\mu\chi\partial_\nu\chi
    -V(\chi)
    \right\}\,,
    \label{eqn:original-action}
\end{align}
where $\xi$ and $q$ are dimensionless parameters, while $M$ and $\chi_0$ are parameters with the mass dimension of one. We note that the $\chi_0$ term in the non-minimal coupling function is introduced to break the $Z_2$ symmetry in the potential so as to avoid producing too large quantum fluctuations at the second inflationary stage~\cite{Braglia:2022phb,Wang:2024vfv}. We expand the scalar potential $V(\chi)$ to the quartic order in $\chi$ such that
\begin{align}
    V(\chi) = V_0 - \frac{1}{2}m^2\chi^2 + \frac{1}{4}\lambda\chi^4\,,
    \label{eqn:original-potential}
\end{align}
where $V_0=m^4/(4\lambda)$ is to make the cosmological constant vanish at the minimum.
The model~\eqref{eqn:original-action} captures many well-explored inflationary models. For instance, the $q = 0$ case corresponds to the standard Higgs-$R^2$ or Starobinsky-Higgs model. On the other hand, the $\xi = 0$ case may be viewed as a simple extension of the Starobinsky model; inflationary analysis for this setup has been discussed in, {\it e.g.}, Refs.~\cite{Huang:2013hsb,Asaka:2015vza,Cheong:2020rao}.
Various other extensions have also been studied in, for instance, Refs.~\cite{Sebastiani:2013eqa,Myrzakulov:2014hca,Myrzakulov:2016tsz,Odintsov:2019evb,Odintsov:2020ilr,Odintsov:2020thl,Odintsov:2020nwm,Odintsov:2021wjz}.
The action~\eqref{eqn:original-action} can thus be regarded as a natural extension of the Higgs-$R^2$ model. Possible effects of the $R^3$ term have been discussed in, {\it e.g.}, Refs.~\cite{Pi:2017gih,Lee:2023wdm,Wang:2024vfv}.

It is well established that $f(R)$ gravity is equivalent to scalar-tensor theory with a scalar degree of freedom~\cite{DeFelice:2010aj,Nojiri:2010wj,Nojiri:2017ncd}. Introducing an auxiliary field $\psi$, the action~\eqref{eqn:original-action} may be written as
\begin{align}
    S = \int d^4x \, \sqrt{-g} \, \bigg\{
    \frac{M_{\rm P}^2}{2}\left[
    1 - \xi\frac{(\chi-\chi_0)^2}{M_{\rm P}^2}
    + \frac{\psi}{3M^2}
    + q \frac{\psi^2}{M^4}
    \right]R
    -\frac{1}{2}g^{\mu\nu}\partial_\mu\chi\partial_\nu\chi
    -V(\chi)
    -U(\psi)
    \bigg\}\,,
    \label{eqn:Jordan-action}
\end{align}
where
\begin{align}
    U(\psi) = \frac{M_{\rm P}^2\psi^2}{12M^2}\left(
    1+\frac{4q}{M^2}\psi
    \right)\,.
    \label{eqn:Jordan-potential-psi}
\end{align}
It is straightforward to show that varying the action~\eqref{eqn:Jordan-action} with respect to $\psi$ gives $\psi = R$ which allows us to recover the original action~\eqref{eqn:original-action}. The scalar field $\psi$ is often called the scalaron. One may note that the mass dimension of the scalaron here is two.

For the analysis of inflationary dynamics, it is more convenient to work in the Einstein frame. To bring the above action~\eqref{eqn:Jordan-action} to the Einstein frame, we perform the conformal transformation~\cite{Dicke:1961gz,Faraoni:1998qx,Fujii:2003pa,DeFelice:2010aj}
\begin{align}
    g_{\mu\nu} \to g_{{\rm E}\mu\nu} = \Omega^2 g_{\mu\nu}
    \,,\quad
    \Omega^2 = 1 - \xi\frac{(\chi-\chi_0)^2}{M_{\rm P}^2} + \frac{\psi}{3M^2} + q\frac{\psi^2}{M^4}\,.
\end{align}
One then obtains
\begin{align}
    S = \int d^4x \, \sqrt{-g_{\rm E}} \, \left[
    \frac{M_{\rm P}^2}{2}R_{\rm E}
    -\frac{3M_{\rm P}^2}{4\Omega^4}g_{\rm E}^{\mu\nu}\partial_\mu\Omega^2\partial_\nu\Omega^2
    -\frac{1}{2\Omega^2}g_{\rm E}^{\mu\nu}\partial_\mu\chi\partial_\nu\chi
    -\frac{V(\chi)+U(\psi)}{\Omega^4}
    \right]\,.
\end{align}
Canonically normalising the first kinetic term with
\begin{align}
    \phi = \sqrt{\frac{3}{2}}M_{\rm P}\ln\Omega^2\,,
\end{align}
which has the mass dimension of one, the Einstein-frame action can be written as
\begin{align}
    S = \int d^4x \, \sqrt{-g_{\rm E}} \, \left[
    \frac{M_{\rm P}^2}{2}R_{\rm E}
    -\frac{1}{2}g_{\rm E}^{\mu\nu}\partial_\mu\phi\partial_\nu\phi
    -\frac{1}{2}e^{-\sqrt{2/3}\phi/M_{\rm P}}g_{\rm E}^{\mu\nu}\partial_\mu\chi\partial_\nu\chi
    -V_{\rm E}(\phi,\chi)
    \right]\,,
    \label{eqn:Einstein-action}
\end{align}
where the Einstein-frame potential is given by
\begin{align}
    V_{\rm E}(\phi,\chi) = e^{-2\sqrt{2/3}\phi/M_{\rm P}}\big[
    V(\chi) + U(\psi(\phi,\chi))
    \big]\,.
    \label{eqn:Einstein-potential}
\end{align}
We note that\footnote{
In principle, there exist two solutions for $\psi$,
\begin{align*}
    \psi_{\pm}(\phi,\chi) = \frac{M^2}{6q}\left[
    \pm\sqrt{1-36q\left(
    1-\xi\frac{(\chi-\chi_0)^2}{M_{\rm P}^2}
    -e^{\sqrt{2/3}\phi/M_{\rm P}}
    \right)} - 1
    \right]\,.
\end{align*}
We choose $\psi_+$ to correctly recover the standard Higgs-$R^2$ scenario in the vanishing-$q$ limit.
}
\begin{align}
    \psi(\phi,\chi) = \frac{M^2}{6q}\left[
    \sqrt{1-36q\left(
    1-\xi\frac{(\chi-\chi_0)^2}{M_{\rm P}^2}
    -e^{\sqrt{2/3}\phi/M_{\rm P}}
    \right)} - 1
    \right]\,.
\end{align}
The cubic term of the Ricci scalar may naturally be viewed as a sub-leading term to the $R^2$ and the non-minimal coupling terms. It is thus desirable to take the $|q| \ll 1$ limit. Assuming that the $q$ term is much less than unity, {\it i.e.,} $36|q(1-\xi(\chi-\chi_0)^2/M_{\rm P}^2-\exp(\sqrt{2/3}\phi/M_{\rm P}))| \ll 1$, we may expand $\psi$ and $U$ up to the leading order in $q$ as
\begin{align}
    \psi(\phi,\chi) \approx
    3M^2\left[
    \xi\frac{(\chi-\chi_0)^2}{M_{\rm P}^2} + e^{\sqrt{2/3}\phi/M_{\rm P}} - 1
    \right]
    -27M^2 q \left[
    \xi\frac{(\chi-\chi_0)^2}{M_{\rm P}^2} + e^{\sqrt{2/3}\phi/M_{\rm P}} - 1
    \right]^2
    \,,
\end{align}
and
\begin{align}
    U \approx 
    \frac{3}{4}M^2M_{\rm P}^2\left[
    \xi\frac{(\chi-\chi_0)^2}{M_{\rm P}^2} + e^{\sqrt{2/3}\phi/M_{\rm P}} - 1
    \right]^2
    -\frac{9}{2}M^2M_{\rm P}^2 q \left[
    \xi\frac{(\chi-\chi_0)^2}{M_{\rm P}^2} + e^{\sqrt{2/3}\phi/M_{\rm P}} - 1
    \right]^3
    \,.
\end{align}
We see that the action reduces to that of the Higgs-$R^2$ model in the $q \to 0$ limit~\cite{Wang:2024vfv}. Throughout the paper, we focus on the regime where $36|q(1-\xi(\chi-\chi_0)^2/M_{\rm P}^2-\exp(\sqrt{2/3}\phi/M_{\rm P}))| \ll 1$ holds, which we call the small-$q$ limit.

\begin{figure}[ht]
	\centering
	\includegraphics[width=0.65\textwidth]{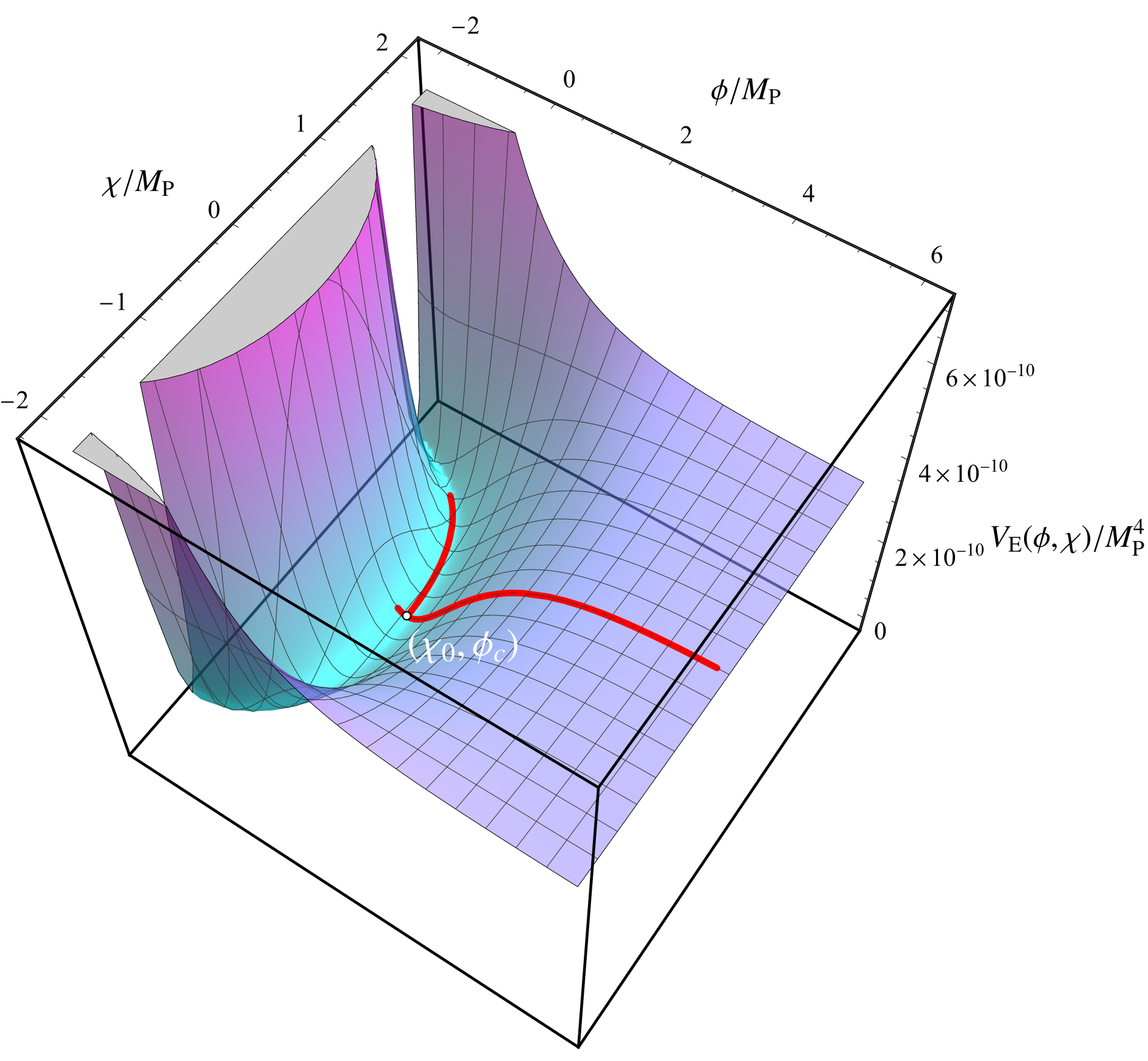}
	\caption{The shape of the potential in the Einstein frame with the parameter choice of $\{M, m, \xi, A, \chi_0, q\} = \{2.2\times 10^{-5}M_{\rm P}, 6.6\times 10^{-6}M_{\rm P}, 0.3125, 1.5, 0.062M, 2.8\times 10^{-5}\}$. The background trajectory is also shown in red.}
	\label{fig:potential}
\end{figure}

As advertised in the introduction, our particular interest is in the two-step inflationary scenario where inflation first occurs along the $\phi$-field direction with the $\chi$ field rapidly rolling down to $\chi_0$, the minimum value of $\chi$ in the large-$\phi$ limit. Crudely speaking, once $\phi$ reaches its minimum, inflation occurs again, but this time, along the $\chi$-field direction until the field hits the vacuum expectation value of $\chi\approx\sqrt{m^2/\lambda}$.
Schematically, inflation comprises three stages. The first stage happens in the large-$\phi$ limit such that $\Omega^2 \gg 1$. Effectively, single-field inflation along the $\phi$-field direction is realised. When the $\phi$ field reaches its local minimum, denoted by $\phi_c$, the first phase of inflation ends. Considering $\phi_c \ll M_{\rm P}$, $\Omega^2$ quickly approaches unity, and the Hubble parameter rapidly drops as well. During this intermediate, second stage, the effective mass-squared of the $\chi$ field becomes negative, setting the stage for the second phase of inflation along the $\chi$-field direction. During the final, third stage, the $\phi$ field settles down to $\phi_c$, and the second phase of inflation along the $\chi$-field direction starts. It is worth noting that $\phi_c$ is not a constant, but a function of the $\chi$ field, $\phi_c(\chi)$, as we shall discuss below. Inflation would then end when the fields reach the global potential minimum, $(\phi_{\rm mim}, \chi_{\rm mim})$, where
\begin{align}
    \phi_{\rm min} \approx \sqrt{\frac{3}{2}}M_{\rm P}\ln\left[
    1-\frac{\xi(m - \sqrt{\lambda}\chi_0)^2}{\lambda M_{\rm P}^2}
    \right]
    \,,\quad
    \chi_{\rm min} \approx \frac{m}{\sqrt{\lambda}}\,.
\end{align}
In Fig.~\ref{fig:potential}, we present the potential shape in the Einstein frame and the background trajectory for a particular choice of the model parameters. Such a two-step inflationary scenario has been extensively studied in, for instance, Refs.~\cite{Pi:2017gih,Cheong:2022gfc,Wang:2024vfv,Wang:2025lti}. In the current work, we aim to scrutinise possible effects of the $R^3$ term.
Closely following Ref.~\cite{Wang:2024vfv}, we present analytical and numerical studies of the inflationary dynamics, highlighting the effects of the $R^3$ term.

\subsection{Background Dynamics}
\label{subsec:backgrounds}
The Einstein-frame action~\eqref{eqn:Einstein-action} takes the form
\begin{align}
    S = \int d^4x \, \sqrt{-g} \, \left[
    \frac{M_{\rm P}^2}{2}R
    -\frac{1}{2}(\partial\phi)^2
    -\frac{1}{2}e^{2b(\phi)}(\partial\chi)^2
    -V_{\rm E}(\phi,\chi)
    \right]\,,
\end{align}
where we have dropped the subscript `E' for notational brevity, except for the potential to avoid any possible confusion with the original $\chi$-field potential~\eqref{eqn:original-potential}.
The background dynamics is governed by the following equations of motion (see, {\it e.g.}, Refs.~\cite{Garcia-Bellido:1995him,Garcia-Bellido:1995hsq,Starobinsky:2001xq,DiMarco:2002eb,DiMarco:2005nq,Choi:2007su,Kim:2013ehu}):
\begin{align}
    H^2 &= \frac{1}{3M_{\rm P}^2} \left(
    \frac{1}{2}\dot{\phi}^2 + \frac{1}{2}e^{2b}\dot{\chi}^2 + V_{\rm E}
    \right)
    \,,\label{eqn:eomH}\\
    0 &= \ddot{\phi} + 3H\dot{\phi} + V_{{\rm E},\phi} - b_{,\phi}e^{2b}\dot{\chi}^2
    \,,\label{eqn:eomphi}\\
    0 &= \ddot{\chi} + \left(3H + 2b_{,\phi}\dot{\phi}\right)\dot{\chi} + e^{-2b}V_{{\rm E},\chi}
    \,,\label{eqn:eomchi}
\end{align}
where a dot denotes the cosmic time derivative, {\it e.g.}, $\dot{\phi}\equiv d\phi/dt$, and the comma denotes the derivative with respect to the field, {\it e.g.}, $V_{{\rm E},\phi}\equiv \partial V_{\rm E}/\partial\phi$. For the model under consideration, we have
\begin{align}
    b(\phi) = -\frac{1}{2}\sqrt{\frac{2}{3}}\frac{\phi}{M_{\rm P}}\,.
\end{align}
From the background equations, one may read that the first Hubble slow-roll parameter is given by
\begin{align}
    \epsilon \equiv -\frac{\dot{H}}{H^2} = \frac{\dot{\phi}^2 + e^{2b}\dot{\chi}^2}{2M_{\rm P}^2H^2}\,.
\end{align}
It is convenient to express the background equations of motion in terms of the $e$-folding number $n$ defined by $dn = Hdt$. Then, the background equations can be rewritten as
\begin{align}
    H^2 &= \frac{V_{\rm E}/M_{\rm P}^2}{3-\epsilon}
    \,,\label{eqn:bkgeomH}\\
    0 &= \phi'' + (3-\epsilon)\phi' + \frac{V_{{\rm E},\phi}}{H^2} - b_{,\phi}e^{2b}\chi'^2
    \,,\label{eqn:bkgeomphi}\\
    0 &= \chi'' + (3-\epsilon+2b_{,\phi}\phi')\chi' + e^{-2b}\frac{V_{{\rm E},\chi}}{H^2}
    \,,
    \label{eqn:bkgeomchi}
\end{align}
and
\begin{align}
    \epsilon = \frac{\phi'^2 + e^{2b}\chi'^2}{2M_{\rm P}^2}\,,
\end{align}
where the prime denotes the derivative with respect to the number of $e$-folds, {\it e.g.}, $\phi' \equiv d\phi/dn$.

\begin{figure}[t!]
	\centering
	\includegraphics[width=0.75\textwidth]{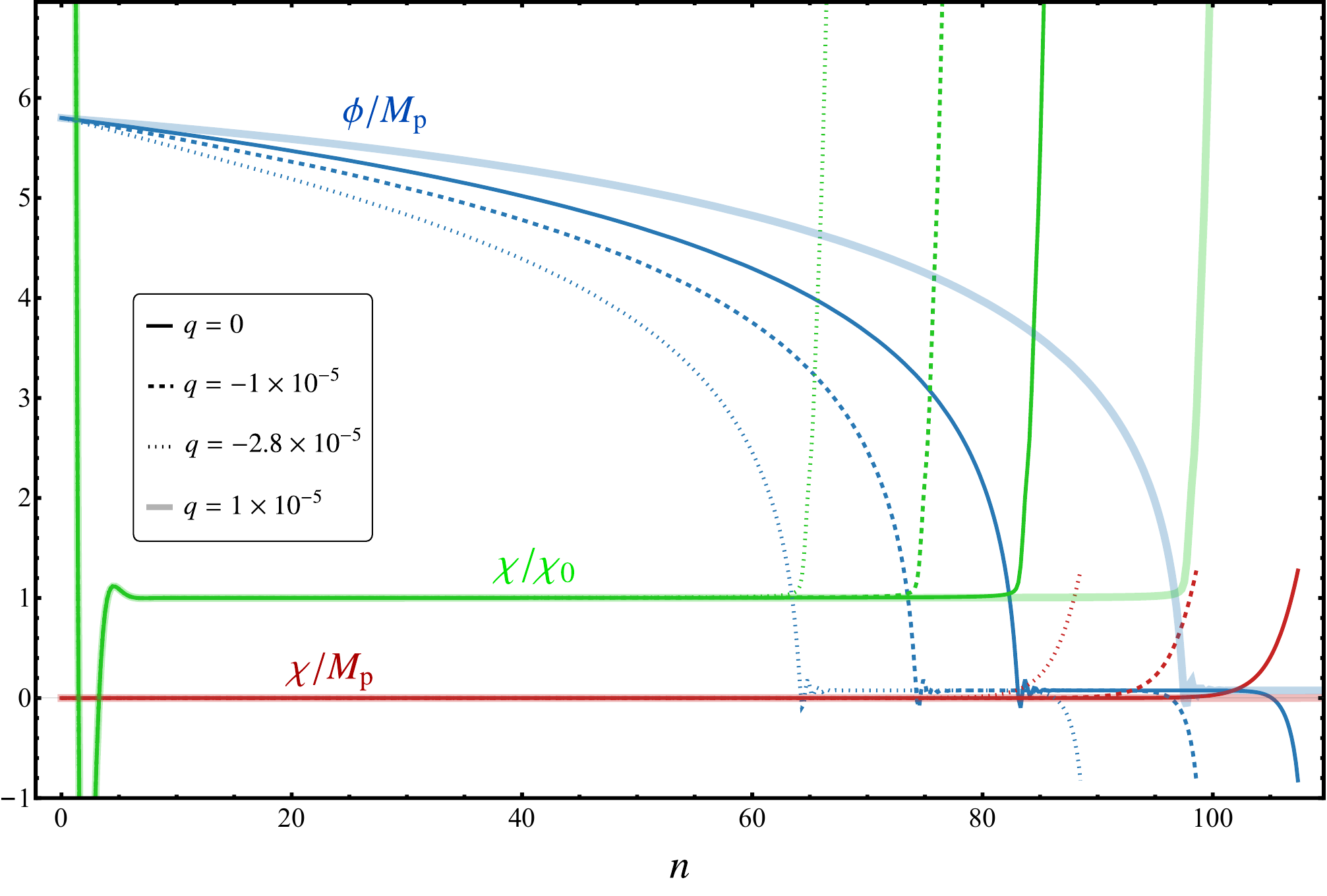}
	\caption{Time evolution of the background $\phi$ field (blue lines) and $\chi$ field (red lines). For clarity, the evolution of $\chi/\chi_0$ is also shown in green. Translucent-thick lines, solid lines, dashed lines, and dotted lines correspond to $q=1 \times 10^{-5}$, $q=0$, $q=-1 \times 10^{-5}$, and $q=-2.8 \times 10^{-5}$, respectively. The rest of the parameters are chosen as $\{M, m, \xi, A, \chi_0\} = \{2.2\times 10^{-5}M_{\rm P}, 6.6\times 10^{-6}M_{\rm P}, 0.3125, 1.5, 0.062M\}$ and remain unchanged in order to show the effect of the $R^3$ term.}
	\label{fig:bgcompare}
\end{figure}

Figure~\ref{fig:bgcompare} shows numerical solutions of the background equations \eqref{eqn:bkgeomH}, \eqref{eqn:bkgeomphi}, and \eqref{eqn:bkgeomchi}. Blue and red lines respectively represent the evolutions of the $\phi$ and $\chi$ fields in terms of the number of $e$-folds $n$, while green lines indicate the evolution of the ratio $\chi/\chi_0$.
In order to clearly show the effect of the $R^3$ term, we change the coefficient of the $R^3$ term, namely $q$, and fix the rest of the parameters as $\{M, m, \xi, A, \chi_0\} = \{2.2\times 10^{-5}M_{\rm P}, 6.6\times 10^{-6}M_{\rm P}, 0.3125, 1.5, 0.062M\}$. Four values of $q$ are chosen: $q=1 \times 10^{-5}$ (translucent-thick lines), $q=0$ (solid lines), $q=-1 \times 10^{-5}$ (dashed lines), and $q=-2.8 \times 10^{-5}$ (dotted lines).
One may see that the first stage of inflation is mainly driven by the $\phi$ field, followed by the second stage where the $\phi$ field oscillates. Afterwards, the $\chi$ field drives the third stage of inflation. Compared to the standard Starobinsky-Higgs model, {\it i.e.}, the $q=0$ case, negative coefficients $q<0$ make the $\phi$ field drop quickly, while positive coefficients $q>0$ delay the $\phi$ field evolution. Below, we analytically study each stage and discuss the effect of the $R^3$ term systematically.

\subsubsection{Stage 1}
During the first stage of inflation, inflation occurs along the $\phi$-field direction when $\phi$ rolls down from a large initial value. Hence, during this stage, we take the large-$\phi$ limit $\phi/M_{\rm P} \gg 1$. Under the slow-roll approximation, the background system can be approximated as
\begin{align}
    \phi' \approx -M_{\rm P}^2\frac{V_{{\rm E},\phi}}{V_{\rm E}}\,.
\end{align}
In the large-$\phi$ limit, from $V_{{\rm E},\chi} = 0$ and $V_{{\rm E},\chi\chi} > 0$, we see that $\chi = \chi_0$ is the minimum of the $\chi$ field. Requiring the effective mass of the $\chi$ field to be larger than the Hubble parameter, so that the $\chi$ field quickly settles down to $\chi_0$, puts a constraint on the non-minimal coupling $\xi$~\cite{Wang:2024vfv}, as will be shown later. In this regime, it is safe to set $\chi = \chi_0$ during the first stage.
Computing $V_{{\rm E},\phi}/V_{\rm E}$ and taking the small-$q$ limit, we obtain, from the background equation of motion, that
\begin{align}
    \frac{\phi'}{M_{\rm P}} \approx
    -2\sqrt{\frac{2}{3}}\frac{(-1+F)-\mu^{-2}}{(-1+F)^2+\mu^{-2}}
    +2\sqrt{6} q F(-1+F)^2\frac{(-1+F)^2+3\mu^{-2}}{[(-1+F)^2+\mu^{-2}]^2}
    \,,
\end{align}
where $F\equiv e^{-2b} = e^{\sqrt{2/3}\phi/M_{\rm P}}$, and, following Ref.~\cite{Wang:2024vfv}, we have defined
\begin{align}
    \mu^2 \equiv \frac{H^2|_{\text{Stage 1}}}{H^2|_{\text{Stage 3}}} \approx \frac{3M^2M_{\rm P}^2}{4V_m}
    \,,\label{eqn:mudef}
\end{align}
with $H|_{\text{Stage 1}}$ ($H|_{\text{Stage 3}}$) being the Hubble parameter during the first (third) stage of inflation, and
\begin{align}
    V_m \equiv V_0 - \frac{1}{2}m^2\chi_0^2 + \frac{1}{4}\lambda\chi_0^4
    \,.\label{eqn:Vmdef}
\end{align}
Since we are considering the case where inflation temporarily halts, we take $\mu^2 \gg 1$.
Hence, under the large-$\phi$ limit and $\mu^{-2} \ll 1$, we obtain
\begin{align}
    \frac{\phi'}{M_{\rm P}} \approx -2\sqrt{\frac{2}{3}}\frac{1}{F} + 2\sqrt{6} q F\,,
\end{align}
which can equivalently be expressed as
\begin{align}
    F' \approx -\frac{4}{3} + 4 q F^2\,.
    \label{eqn:FEOMapp}
\end{align}
We see that at the leading order $F'\approx -4/3$, which indicates that $F$ decreases when $\phi$ rolls down during the first stage of inflation. We also see that the contribution of the $R^3$ term to $F$ is positive (negative) for $q > 0$ ($q < 0$). Solving Eq.~\eqref{eqn:FEOMapp} up to the $\mathcal{O}(q)$ order, we find
\begin{align}
	F = F_0 - \frac{4}{3}(n-n_0)
	+q F_0^3 \left\{
	1-\left[
	1-\frac{4}{3F_0}\left(n-n_0\right)
	\right]^3
	\right\}\,,
\end{align}
where $F_0$ is the value of $F$ at $n=n_0$, {\it i.e.}, the initial value, which can be found by specifying the end of the first stage of inflation.
The first stage ends when $\epsilon \approx \phi^{\prime 2}/(2M_{\rm P}^2)|_{\phi=\phi_{*1}} = 1$. From the background equation of motion, we find that the value of $F$ at $\phi=\phi_{*1}$ is given by
\begin{align}
    F_{*1} &= 1 + \frac{1}{\sqrt{3}} + \frac{1}{\sqrt{3}}S
    \nonumber\\&\quad
    +q \left[
    -4(2+\sqrt{3})(1+S)
    +\frac{36+23\sqrt{3}+3(4+3\sqrt{3})S}{\mu^2}
    +\frac{3(-1+S+\sqrt{3}S)}{\mu^4}
    \right]
    \nonumber\\&\quad\quad\times
    \left[
    \sqrt{3}(1+S) - \frac{3(2+\sqrt{3}+S)}{\mu^2}
    \right]^{-1}
    \,,
\end{align}
up to $\mathcal{O}(q)$ in the small-$q$ limit, where
\begin{align}
    S \equiv \sqrt{1-\frac{3+2\sqrt{3}}{\mu^2}}\,.
\end{align}
From the equation above, it immediately follows that $\mu^2 \geq 3+2\sqrt{3}$, which is satisfied in our case for $\mu^2 \gg 1$. In terms of $\phi$,
\begin{align}
    \phi_{*1} &= \sqrt{\frac{3}{2}}M_{\rm P}\ln\left(
    1+\frac{1}{\sqrt{3}}+\frac{1}{\sqrt{3}}S
    \right)
    \nonumber\\&\quad
    -3\sqrt{\frac{3}{2}} q M_{\rm P} 
    \left\{
    (3+\sqrt{3}+\sqrt{3}S)\left[\sqrt{3}(1+S)-\frac{3(2+\sqrt{3}+S)}{\mu^2}\right]
    \right\}^{-1}
    \label{eqn:phis1}\\&\quad\quad\times
    \left[
    4(2+\sqrt{3})(1+S)
    -\frac{36+23\sqrt{3}+3(4+3\sqrt{3})S}{\mu^2}
    +\frac{3-3(1+\sqrt{3})S}{\mu^4}
    \right]
    \,.\nonumber
\end{align}
In the $\mu^2\gg 1$ limit, we have
\begin{align}
    \phi_{*1} \approx 0.94 M_{\rm P} - 4.9 q M_{\rm P}\,.
    \label{eqn:phis1app}
\end{align}
We see that the presence of the $R^3$ term makes the value of $\phi$ at the end of the first stage smaller (larger) for $q > 0$ ($q < 0$).

With the value of $F$ (or $\phi$) at the end of the first stage, we can fix $F_0$, which is, up to $\mathcal{O}(q)$, given by
\begin{align}
    F_0 =
    F_{*1} + \frac{4}{3}(n_{*1} - n_0)
    -\frac{4}{27} q (n_{*1} - n_0)\left[
    27F_{*1}^2 + 36F_{*1}(n_{*1} - n_0)
    +16(n_{*1}-n_0)^2
    \right]\,.
\end{align}
Therefore, up to $\mathcal{O}(q)$, we obtain
\begin{align}
    F = F_{*1} + \frac{4}{3}(n_{*1}-n)
    -\frac{4}{27} q (n_{*1}-n)\left[
    27F_{*1}^2 + 36 F_{*1}(n_{*1}-n) + 16(n_{*1}-n)^2
    \right]
    \,.
\end{align}

As the first stage ends, $\phi$ approaches its local potential minimum $\phi_c$. Recall that $\chi$ has been trapped at $\chi \approx \chi_0$.
The value of $\phi_c$ can be found by looking at $V_{{\rm E},\phi}$ and $V_{{\rm E},\phi\phi}$ at $\phi=\phi_c$ and $\chi=\chi_0$:
\begin{align}
    \phi_c = \sqrt{\frac{3}{2}}M_{\rm P}\left[
    \ln\left(1+\mu^{-2}\right)
    +3q \mu^{-4}\left(\frac{3+\mu^{-2}}{1+\mu^{-2}}\right)
    \right]\,.
\end{align}
In the $\mu^2 \gg 1$ limit,
\begin{align}
    \phi_c \approx \sqrt{\frac{3}{2}}M_{\rm P}\mu^{-2}
    -\frac{1}{2}\sqrt{\frac{3}{2}}M_{\rm P} \mu^{-4}(1-18q) \ll M_{\rm P}
    \,.\label{eqn:smallness-of-phic}
\end{align}
We see that the $q$ correction enters at the next leading order.

To discuss the dynamics of the $\chi$ field, let us first examine its effective mass-squared. As $\chi$ is very close to $\chi_0$, we define $\Delta \chi \equiv \chi-\chi_0$ with $|\Delta \chi|/\chi_0 \ll 1$, which satisfies
\begin{align}
    &\ddot{\Delta \chi} + \left(
    3H+2b_{,\phi}\dot{\phi}
    \right)\dot{\Delta \chi} + m_{\chi,{\rm eff}}^2\Delta\chi = 0\,,\\
    &m_{\chi,{\rm eff}}^2\equiv e^{-2b}V_{{\rm E},\chi\chi}\,.
\end{align}
At the beginning of the first stage, as $F\gg 1$, the effective mass-squared is given by
\begin{align}
    m_{\chi,{\rm eff}}^2\big|_{F\gg1} \approx
    3M^2\xi\left(
    1 - 9q e^{\sqrt{\frac{2}{3}}\frac{\phi}{M_{\rm P}}}
    \right)\,,
\end{align}
up to $\mathcal{O}(q)$.
As we discussed above, we require that $m_{\chi,{\rm eff}}^2 \gg H^2$, which puts the bound of $\xi \gg 1/12$ on the non-minimal coupling; see also Ref.~\cite{Wang:2024vfv}.
We note further that the requirement of $m_{\chi,{\rm eff}}^2 > 0$ during the first stage poses an additional constraint that $q$ cannot be positively large, $q < e^{-\sqrt{2/3}\phi/M_{\rm P}}/9$.

\subsubsection{Stage 2}
At the start of the intermediate, second stage, or, equivalently, at the end of the first stage, $\phi=\phi_{*1}$ and $\chi=\chi_0$ so that 
\begin{align}
    m_{\chi,{\rm eff}}^2\big|_{(\phi_{*1}, \chi_0)} &=
    \frac{3\left[3\lambda \chi_0^2-m^2-\sqrt{3}M^2\xi(1+S)\right]}{3+\sqrt{3}(1+S)}
    \nonumber\\&\quad
    +\frac{9q}{[3+\sqrt{3}(1+S)]^2}
    \bigg\{
    \frac{m^2+3\left(M^2\xi-\lambda \chi_0^2\right)}{\sqrt{3}(1+S)-3(2+\sqrt{3}+S)/\mu^2}
    \nonumber\\&\quad\quad\times
    \left[
    -4(2+\sqrt{3})(1+S)
    +\frac{36+23\sqrt{3}+3(4+3\sqrt{3})S}{\mu^2}
    +\frac{3(-1+S+\sqrt{3}S)}{\mu^4}
    \right]
    \nonumber\\&\quad\quad
    -3M^2\xi[3+\sqrt{3}(1+S)](1+S)^2
    \bigg\}\,,
\end{align}
up to $\mathcal{O}(q)$.
We note that $m_{\chi,{\rm eff}}^2>0$ should be satisfied at the start of the second stage. In the absence of the $q$ contribution, this condition reduces to the following form:
\begin{align}
    \xi >\frac{m^2 - 3\lambda \chi_0^2}{\sqrt3M^2}\left(
    1+\sqrt{1-\frac{3+2\sqrt{3}}{\mu^2}}
    \right)^{-1}\approx\frac{m^2 - 3\lambda \chi_0^2}{2\sqrt3M^2}\,,
    \label{eqn:xicon1}
\end{align}
where we have taken the $\mu^2 \gg 1$ limit in the last step.

At the end of second stage, on the other hand, $m_{\chi,{\rm eff}}^2 < 0$ needs to be met so that $\chi$ may start to roll down, causing the second phase of inflation, {\it i.e.}, the third stage. Hence, at $\phi=\phi_c$ and $\chi=\chi_0$,
\begin{align}
    m_{\chi,{\rm eff}}^2\big|_{\left(\phi_c, \chi_0\right)} = \frac{3M^2\xi-(m^2-3\lambda\chi_0^2)\mu^2}{1+\mu^2} + 3q\frac{(m^2-3\lambda\chi_0^2)(1+3\mu^2)-6M^2\xi}{\mu^2(1+\mu^2)^2} + \mathcal{O}(q^2)<0
\end{align}
At the leading order, this implies that
\begin{align}
    \xi < \frac{\mu^2}{3M^2}\left(m^2-3\lambda\chi_0^2\right)\,.
    \label{eqn:xicon2}
\end{align}

Therefore, the combination of the conditions \eqref{eqn:xicon1} and \eqref{eqn:xicon2} gives a constraint on the value of $\xi$ for the inflationary trajectory to have a turn. Under this constraint, we now solve the background system to examine the evolution dynamics during the second stage. Note that during this stage, $\chi=\chi_0$ while $\phi$ evolves from $\phi_{*1}$ to $\phi_c$. The background equation of motion for $\phi$ is
\begin{align}
    \ddot{\phi} + 3H\dot{\phi} + V_{{\rm E},\phi} = 0\,.
\end{align}
Setting $\Delta \phi \equiv \phi-\phi_c$, the background equation for $\Delta\phi$ is expressed as
\begin{align}
    0 &= \ddot{\Delta\phi} + 3H\dot{\Delta\phi} + m_{\phi,{\rm eff}}^2\Delta\phi + \mathcal{O}(q^2,\Delta\phi^2)
    \,,\label{eqn:EOMdeltaphi}\\
    m_{\phi,{\rm eff}}^2&\equiv M^2\left[
    \frac{\mu^2}{1+\mu^2} - 6q \frac{2+6\mu^2+3\mu^4}{\mu^2(1+\mu^2)^2}
    \right]
    \,.\label{eqn:mphieff2}
\end{align}
We note that the $q$ term in $m_{\phi,{\rm eff}}^2$ is suppressed by a factor of $\mu^{-2}$. In order to obtain the approximate solution for $\Delta\phi$, let us compute the Hubble parameter and compare it with $m_{\phi,{\rm eff}}^2$. At $\phi=\phi_{*1}$ and $\chi=\chi_0$, if we insert Eqs.~\eqref{eqn:Vmdef} and \eqref{eqn:phis1} into $H^2 \approx V_{\rm E}/(3M_{\rm P}^2)$, we find that
\begin{align}
    H^2 \approx 0.07 M^2\,,
\end{align}
where we have taken the small-$q$ limit as well as the large-$\mu^2$ limit.
On the other hand, from Eq.~\eqref{eqn:mphieff2}, for $\mu^2\gg1$, we have $m_{\phi,{\rm eff}}^2\approx M^2$. So we have $m_{\phi,{\rm eff}}^2\gg H^2$ at the start of the second stage.
Note also that
\begin{align}
    H|_{\text{at the end of Stage 2}} < H|_{\text{at the start of Stage 2}}\,,
\end{align}
{\it i.e.}, $H$ decreases in time. 
Hence, during the second stage, we may safely take $m_{\phi,{\rm eff}}^2 \gg H^2$ and neglect the time variation of the Hubble parameter to obtain the approximate solution for $\Delta\phi$:
\begin{align}
    \Delta \phi(t) = \Delta\phi(t_{* 1}) \left(
    \frac{a(t_{* 1})}{a(t)}
    \right)^{3/2}\cos\left[
    m_{\phi,{\rm eff}}(t-t_{*1})
    \right]\,.
\end{align}

The potential near $\phi=\phi_c$ can be approximated as
\begin{align}
    V_{\rm E} &\approx
    \frac{3M^2M_{\rm P}^2}{4\mu^2} - \sqrt{\frac{3}{2}}\frac{M^2M_{\rm P}^2}{\mu^2}\left(\frac{\phi_c}{M_{\rm P}}\right)
    -\sqrt{\frac{3}{2}}\frac{M^2M_{\rm P}^2}{\mu^2}\left(\frac{\Delta\phi}{M_{\rm P}}\right)
    \nonumber\\ &\qquad
    +\frac{1}{2}\frac{M^2M_{\rm P}^2}{\mu^2}(2+\mu^2)\left(\frac{\phi_c}{M_{\rm P}}\right)^2
    +\frac{1}{2}\frac{M^2M_{\rm P}^2}{\mu^2}(2+\mu^2)\left(\frac{\Delta\phi}{M_{\rm P}}\right)^2
    \nonumber\\ &\qquad
    +\frac{M^2M_{\rm P}^2}{\mu^2}(2+\mu^2)\left(\frac{\phi_c}{M_{\rm P}}\right)\left(\frac{\Delta\phi}{M_{\rm P}}\right)
    \nonumber\\ &\approx
    \frac{M^2}{2}\Delta\phi^2 + \frac{3M^2M_{\rm P}^2}{4\mu^2}
    \,,
\end{align}
where we have taken the large-$\mu^2$ limit. We note that the $q$ contribution is suppressed by $\mu^{-4}$.
Thus, the Hubble parameter is given, as $\dot{\chi}\approx 0$, by
\begin{align}
    H^2 \approx
    \frac{1}{3M_{\rm P}^2}\left(
    \frac{1}{2}\dot{\Delta\phi}^2 + V_{\rm E}
    \right)
    \approx
    \frac{M^2}{3M_{\rm P}^2}\left[
    \frac{\Delta\phi_{*1}^2}{2}\left(\frac{a_{*1}}{a}\right)^3
    +\frac{3M_{\rm P}^2}{4\mu^2}
    \right]
    \,,
\end{align}
where we have taken the large-$\mu^2$ limit and $m_{\phi,{\rm eff}}^2 \gg H^2$. In this expression, $\Delta\phi_{*1} \equiv \Delta\phi(t_{*1})$ and $a_{*1}\equiv a(t_{*1})$.
From Eq.~\eqref{eqn:smallness-of-phic}, we see that $\phi_c$ is suppressed by the large factor of $\mu^2$, and thus, we may express $\phi = \Delta\phi + \phi_c \approx \Delta\phi$. Then, we get
\begin{align}
    H^2 \approx \frac{M^2}{3M_{\rm P}^2}\left[
    \frac{\phi_{*1}^2}{2}\left(\frac{a_{*1}}{a}\right)^3
    +\frac{3M_{\rm P}^2}{4\mu^2}
    \right]\,.
\end{align}
The first Hubble slow-roll parameter is then given by
\begin{align}
    \epsilon = 3 - \frac{V_{\rm E}}{H^2M_{\rm P}^2}
    \approx
    \frac{3\phi_{*1}^2(a_{*1}/a)^3\sin^2[M(t-t_{*1})]}{\phi_{*1}^2(a_{*1}/a)^3+3M_{\rm P}^2/(2\mu^2)}\,,
\end{align}
where we have again taken the large-$\mu^2$ limit. Taking the time-average gives
\begin{align}
    \langle \epsilon \rangle \approx
    \frac{3\phi_{*1}^2(a_{*1}/a)^3}{2\phi_{*1}^2(a_{*1}/a)^3+3M_{\rm P}^2/\mu^2}\,.
\end{align}
We define the end of the second stage as the point $t_{*2}$ at which $\langle\epsilon\rangle|_{t_{*2}}=1/p$ where $p>1$ as in Ref.~\cite{Wang:2024vfv}. Then, we read
\begin{align}
    \frac{a_{*2}}{a_{*1}} = \left(
    \frac{3M_{\rm P}^2}{\mu^2\phi_{*1}^2(3p-2)}
    \right)^{-1/3}\,,
\end{align}
where $a_{*2}\equiv a(t_{*2})$. In other words, the duration of the second stage is
\begin{align}
    \Delta n (\text{Stage 2}) = n_{*2} - n_{*1} = -\frac{1}{3}\ln\left[
    \frac{3}{3p-2}\left(\frac{M_{\rm P}}{\mu\phi_{*1}}\right)^2
    \right]
    \approx \frac{1}{3}\ln\mu^2
    \,,
\end{align}
for an appropriate $p$ value. Here, $n_{*2} \equiv n(t_{*2})$. We see that for $\mu^2 \gg 20$, $\Delta n (\text{Stage 2}) \gtrsim \mathcal{O}(1)$.

We stated earlier that during the second stage, the $\chi$ field becomes unstable and gets ready to roll down slowly, while the $\phi$ field begins damped oscillations around its local minimum $\phi_c\approx\sqrt{3/2}M_{\rm P}/\mu^2 \ll M_{\rm P}$. As a consistency check, let us evaluate the value of $\phi$ at the end of the second stage, $\phi_{*2}$:
\begin{align}
    \phi_{*2} = \phi_c + \Delta\phi(t_{*2})
    = \phi_c + \Delta\phi(t_{*1})\left[
    \frac{a(t_{*1})}{a(t_{*2})}
    \right]^{3/2}\cos\left[
    m_{\phi,{\rm eff}}(t_{*2}-t_{*1})
    \right]\,.
\end{align}
Thus, the magnitude of the field value is
\begin{align}
    \phi_{*2} =
    \phi_c + \Delta\phi(t_{*1})\left[
    \frac{a(t_{*1})}{a(t_{*2})}
    \right]^{3/2}
    =
    \phi_c + \frac{M_{\rm P}}{\mu}\sqrt{\frac{3}{3p-2}}\left(
    1-\frac{\phi_c}{\phi_{*1}}
    \right)
    \approx
    \phi_c\,,
\end{align}
which agrees with the expectation.
Hence, we shall use $\phi_{*2}\approx\phi_c$ as the initial value for the $\phi$ field at the beginning of the final (third) stage.
For the $\chi$ field, we take $\chi_{*2}\approx\chi_0$ as the initial value at the onset of the third stage. We stress that, although $\chi_{*2} \approx \chi_0 \approx \chi_{*1}$, {\it i.e.}, $\chi$ essentially stayed unmoved, the effective mass-squared changed its sign.

\subsubsection{Stage 3}
During Stage 3, the inflationary trajectory is mainly along the $\chi$-field direction while oscillating along the $\phi$-field direction around $\phi_c$. It is worth noting that the local minimum $\phi_c$ is a function of $\chi$, and thus, the inflationary trajectory is not a simple, straight line along the $\chi$-field direction.
We can find $\phi_c(\chi)$ from
\begin{align}
    \frac{\partial V_{\rm E}}{\partial\phi}\bigg\vert_{\phi=\phi_c(\chi)} = 0\,.
\end{align}
Up to $\mathcal{O}(q)$, we obtain
\begin{align}
    F(\phi_c(\chi)) &\approx
    1-\xi\frac{(\chi-\chi_0)^2}{M_{\rm P}^2}
    +\mu^{-2}\left(
    1-\xi\frac{(\chi-\chi_0)^2}{M_{\rm P}^2}
    \right)^{-1}\left[
    1-2\left(\frac{\chi}{\chi_g}\right)^2
    +\left(\frac{\chi}{\chi_g}\right)^4
    \right]
    \nonumber\\&\qquad
    +q\Bigg\{
    9\mu^{-4}\left(
    1-\xi\frac{(\chi-\chi_0)^2}{M_{\rm P}^2}
    \right)^{-2}\left[
    1-2\left(\frac{\chi}{\chi_g}\right)^2
    +\left(\frac{\chi}{\chi_g}\right)^4
    \right]^2
    \nonumber\\&\qquad\qquad
    +3\mu^{-6}\left(
    1-\xi\frac{(\chi-\chi_0)^2}{M_{\rm P}^2}
    \right)^{-4}\left[
    1-2\left(\frac{\chi}{\chi_g}\right)^2
    +\left(\frac{\chi}{\chi_g}\right)^4
    \right]^3
    \Bigg\}
    \,,\label{eqn:Fphic}
\end{align}
where $\chi_g\equiv m/\sqrt{\lambda}$, and we have assumed that $\chi_0\ll\chi_g$. 
We see that the $q$ corrections in Eq.~\eqref{eqn:Fphic} are suppressed by $\mu^{-4}$ and $\mu^{-6}$, respectively. In the following, we omit the argument and simply denote $\phi_c(\chi)$ by $\phi_c$.

Let us work out the evolution of $\phi$. Near $\phi_c$, setting
\begin{align}
    \phi = \phi_c + \Delta\phi\,,
\end{align}
the equation of motion for $\phi$ becomes
\begin{align}
    \ddot{\Delta\phi} + 3H\dot{\Delta\phi} + M^2\Delta\phi \approx 0\,,
\end{align}
where we have taken the large-$\mu^2$ limit, $\chi\ll M_{\rm P}$, and $\chi_0 \ll M_{\rm P}$; we note that the $q$ correction is suppressed by $\mu^{-2}$. The equation of motion is thus approximately the same as the one during Stage 2, given in Eq.~\eqref{eqn:EOMdeltaphi}. The solution is similarly expressed as
\begin{align}
    \Delta\phi(t) = \Delta\phi(t_{*2})\left[
    \frac{a(t_{*2})}{a(t)}
    \right]^{3/2}\cos\left[
    M(t-t_{*2})
    \right]\,,
\end{align}
In terms of $n$, since
\begin{align}
    n-n(t_{*2}) \approx H_2(t-t_{*2})\,,
\end{align}
where $H_2\approx M/(2\mu)$ is the Hubble parameter during Stage 2, we obtain
\begin{align}
    \Delta\phi = \Delta\phi_{*2}e^{-\frac{3}{2}(n-n_{*2})}\cos\left[
    2\mu\left(n-n_{*2}\right)
    \right]
    \,.\label{eqn:deltaphiS3}
\end{align}
Here, the subscript `$*2$' denotes that the quantities are evaluated at $t_{*2}$ or, equivalently, $n_{*2}$.

Next, let us analyse the evolution of $\chi$. The equation of motion \eqref{eqn:eomchi} is equivalently expressed as
\begin{align}
    \chi'' + \left(
    3-\epsilon-\frac{\sqrt{2/3}}{M_{\rm P}}\phi'
    \right)\chi' + F\frac{V_{{\rm E},\chi}}{H^2} = 0
    \,.\label{eqn:eomchiS3}
\end{align}
As $\chi$ slowly rolls down the potential at the onset of Stage 3, using the slow-roll approximation for the $\chi$ field together with $\phi \approx \phi_c$, Eq.~\eqref{eqn:eomchiS3} can be approximated as
\begin{align}
    \chi' \approx -\frac{FV_{{\rm E},\chi}}{3H^2}\bigg\vert_{\phi\approx\phi_c}\,.
\end{align}
Taking $\chi\ll\chi_g$ and $\chi_0\ll\chi_g$, up to $\mathcal{O}(q)$, we obtain
\begin{align}
    \chi' \approx 4\xi\left\{
    \left[
    \frac{M_{\rm P}^2}{\xi \chi_g^2}\left(
    1-3q\mu^{-4}
    \right)-1
    \right]\chi + \chi_0
    \right\}
    \,.\label{eqn:chiS3}
\end{align}
It is evident from Eq.~\eqref{eqn:chiS3} that the $q$ correction to $\chi$ during Stage 3 is suppressed by $\mu^{-4}$. The solution of $\chi$ during Stage 3 is obtained as
\begin{align}
    \chi(n) &=
    \chi\left(n_{*2}\right)\exp\left\{
    4\xi\left[
    \frac{M_{\rm P}^2}{\xi\chi_g^2}\left(1-3q\mu^{-4}\right)-1
    \right](n-n_{*2})
    \right\}
    \nonumber\\&\quad
    +\frac{\xi\chi_0\chi_g^2}{M_{\rm P}^2(1-3q\mu^{-4})-\xi\chi_g^2}\left[
    \exp\left\{
    4\xi\left[
    \frac{M_{\rm P}^2}{\xi\chi_g^2}\left(1-3q\mu^{-4}\right)-1
    \right](n-n_{*2})
    \right\}
    -1
    \right]\,,\nonumber\\
    &=
    \frac{\chi_0}{1-3q\mu^{-4}-\xi\chi_g^2/M_{\rm P}^2}
    \nonumber\\&\quad\times
    \left\{
    \exp\left[
    \frac{4-12q\mu^{-4}-4\xi\chi_g^2/M_{\rm P}^2}{\chi_g^2/M_{\rm P}^2}(n-n_{*2})
    \right]\left(1-\frac{3q}{\mu^4}\right)
    -\frac{\xi\chi_g^2}{M_{\rm P}^2}
    \right\}
    \,.\label{eqn:chisolS3}
\end{align}
where we have taken $\chi(n_{*2})=\chi_0$ in the last step. We note that a necessary condition for the slow-roll of $\chi$ is that the magnitude of the coefficient in the exponent of Eq.~\eqref{eqn:chisolS3} should be small, {\it i.e.},
\begin{align}
    \Bigg\vert
    4\left(\frac{M_{\rm P}}{\chi_g}\right)^2\left[
    1-3q\mu^{-4} - \xi\left(\frac{\chi_g}{M_{\rm P}}\right)^2
    \right]
    \Bigg\vert \ll 1\,,
    \label{eqn:chi-SR-condition}
\end{align}
Neglecting the small $q$ correction, the condition \eqref{eqn:chi-SR-condition} can be expressed as a constraint on $\xi$ as
\begin{align}
    \frac{\lambda M_{\rm P}^2}{m^2} - \frac{1}{4}\ll
    \xi\ll
    \frac{\lambda M_{\rm P}^2}{m^2} + \frac{1}{4}
    \,,\label{eqn:chi-SR-condition-xi}
\end{align}
where we have used $\chi_g=m/\sqrt{\lambda}$.
The slow-roll inflation along $\chi$ ends when
\begin{align}
    \epsilon = \frac{1}{2M_{\rm P}^2}\left(
    \phi^{\prime 2} + F^{-1}\chi^{\prime 2}
    \right) \approx \frac{\chi^{\prime 2}}{2M_{\rm P}^2} = \mathcal{O}(1)\,.
\end{align}
Using the solution \eqref{eqn:chisolS3} for $\chi$, we obtain
\begin{align}
    \Delta n (\text{Stage 3}) &= n_{*3} - n_{*2}
    \nonumber\\&=
    \frac{1}{4}\left(\frac{\chi_g}{M_{\rm P}}\right)^2
    \left(1-\frac{3q}{\mu^4}-\xi\frac{\chi_g^2}{M_{\rm P}^2}\right)^{-1}
    \ln\left[
    \frac{1}{2\sqrt{2}}\frac{\chi_g^2}{M_{\rm P}\chi_0}\left(
    1-\frac{3q}{\mu^4}
    \right)^{-1}
    \right]\,.
\end{align}
Hence, the value of $\chi$ at $n_{*3}$ can be expressed as
\begin{align}
    \chi_{*3}\equiv\chi(n_{*3}) = \frac{\chi_0-\frac{\sqrt{2}M_{\rm P}}{4\xi}}{1-(1-3q\mu^{-4})\frac{M_{\rm P}^2}{\xi\chi_g^2}}\,.
    \label{eqn:chistar3}
\end{align}
We note that, from the constraint $\chi_0 < \chi_{*3} < \chi_g$, we get, ignoring the negligible $q$ term,
\begin{align}
    \chi_g\left[
    1+\frac{M_{\rm P}^2}{\xi\chi_g^2}\left(
    \frac{\sqrt{2}\chi_g}{4M_{\rm P}} - 1
    \right)
    \right] < \chi_0 < \frac{\sqrt{2}\chi_g^2}{4M_{\rm P}}\,.
\end{align}
We note that, when $\sqrt{2}\chi_g < 4M_{\rm P}$, if $\xi\chi_g^2 < M_{\rm P}^2$, the above condition becomes satisfied.

To summarise, from the analytical slow-roll analysis presented in this section, it is evident that the presence of the $R^3$ term most strongly affects Stage 1 of inflation and becomes less effective in later stages. This observation already indicates that the CMB scale is to be most heavily affected by the $R^3$ term. The spectral index $n_s$ and the tensor-to-scalar ratio $r$ will thus be modified due to the inclusion of the $R^3$ term. In the following, we will show this systematically by studying the cosmological perturbations in this model.

\subsection{Cosmological Perturbations}
\label{subsec:perturbations}
We now discuss perturbations and compute the curvature power spectrum. While one can solve the perturbed system numerically without any approximations, we may grasp an analytical understanding by adopting the $\delta N$ formalism. The $\delta N$ formalism relates the curvature perturbation to the difference of the number of $e$-folds between an initial flat hypersurface and a final uniform energy density hypersurface; see, {\it e.g.}, Ref.~\cite{Sasaki:1995aw}. The initial flat hypersurface is taken to be at the horizon-crossing time, denoted by $n_k$, and we take the final time to be $n_c$ after which the background trajectories converge. For $n>n_c$, the curvature perturbation will then remain conserved. 
It is pointed out in Ref.~\cite{Wang:2024vfv} that isocurvature perturbations remain negligible during Stage 1 and Stage 2, while they get enhanced during Stage 3, which in turn source the curvature perturbations. 
The isocurvature perturbations then decay away rapidly before the end of inflation. 
Based on this result, in this work, we choose $n_c$ to be the end-of-inflation point, {\it i.e.}, $n_c=n_{*3}$.

\begin{figure}[t!]
	\centering
	\includegraphics[width=0.8\textwidth]{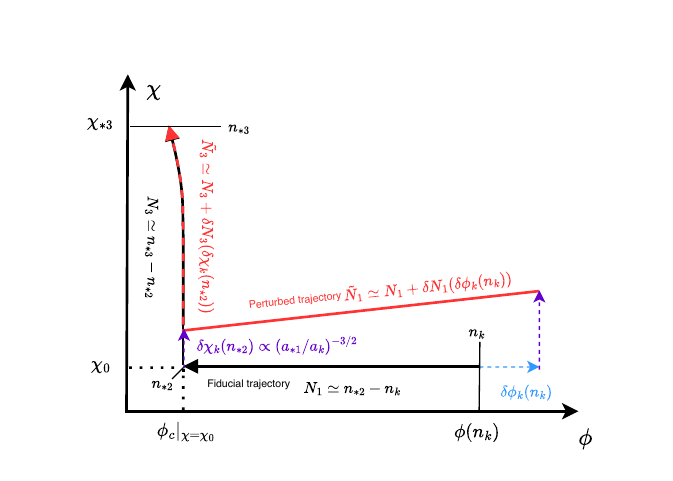}
	\caption{A schematic diagram for the $\delta N$ formalism for modes that exit the horizon slightly before the end of Stage 1. The black solid arrows represent the fiducial background solution on the phase diagram, and the red solid arrows represent the solution after perturbing the initial condition. The purple and blue dotted arrows show the tendency of the change in the solution induced by perturbing $\chi$ by $\delta\chi_k$ and $\phi$ by $\delta\phi_k$, respectively. The end of inflation is defined on the comoving slicing  $\chi=\chi_{*3}$.}
	\label{fig:deltaN}
\end{figure}

\begin{figure}[t!]
	\centering
	\includegraphics[width=0.8\textwidth]{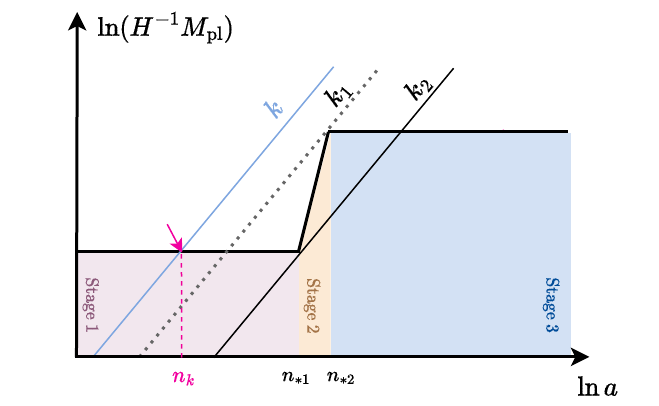}
	\caption{Diagram of the horizon evolution for inflation with a break. The black dotted (solid) line shows the evolution of the proper wavelength of the $k_1$ ($k_2$) mode. The blue solid line indicates the mode of interest which exits the horizon during the first stage of inflation at $n_k$. }
	\label{fig:kposition}
\end{figure}

According to the $\delta N$ formalism, the curvature perturbation is given by
\begin{align}
    \mathcal{R}(n_c) \approx \delta N(n_c, \varphi^a(n_k))
    = N(n_c,\varphi^a(n_k)+\delta\varphi^a(n_k))
    -N(n_c,\varphi^a(n_k))\,,
\end{align}
where $\varphi^a = \{\phi,\chi\}$ in our case, $\delta\varphi^a$ denote the field fluctuations, and $N(n_c,\varphi^a(n_k))$ denotes the number of $e$-folds from $n_k$ to $n_c$. A schematic diagram for the $\delta N$ formalism in our model is illustrated in Fig.~\ref{fig:deltaN}.
Up to the first order in $\delta\varphi^a$, which is sufficient for the discussion of the curvature power spectrum, we find that
\begin{align}
    \mathcal{R}(n_c) \approx \frac{\partial N}{\partial \varphi^a}\Bigg\vert_{n_k} \delta\varphi^a(n_k)\,.
\end{align}
In our case, there exist three stages of inflation. Thus, we may express $\delta N$ as follows, depending on the mode of interest:
\begin{align}
    \mathcal{R}(n_c)\approx
    \begin{cases}
    \delta N(n_{*1}, \varphi^a(n_k))
    +\delta N(n_{*2}, \varphi^a(n_{*1}))
    +\delta N(n_{*3}, \varphi^a(n_{*2})) & \text{  for  }n_k < n_{*1}\,,
    \\
    \delta N(n_{*2}, \varphi^a(n_k))
    +\delta N(n_{*3}, \varphi^a(n_{*2})) & \text{  for  }n_{*1} < n_k < n_{*2}\,,
    \\
    \delta N(n_{*3}, \varphi^a(n_k)) & \text{  for  }n_k > n_{*2}\,.
    \end{cases}
\end{align}
For instance, if a mode $k$ exits the horizon during the first stage, we need to consider the contributions from all three stages, while if a mode $k$ exits during the third stage, only the contribution from the third stage is needed; see Fig.~\ref{fig:kposition} for a schematic diagram.

Since the duration of the intermediate, second stage is rather short, we may safely assume that the mode that touches the horizon at the beginning of the Stage 2 (and exits the horizon during Stage 3), say $k_2$, and the mode that touches the horizon at the end of Stage 2 (and exits during Stage 1), say $k_1$, are approximately identical. In other words, we may consider the following two regions: (i) horizon crossing during the first stage ($k < k_1$), and (ii) horizon crossing during the third stage ($k > k_1$).
Under the above assumption, we have
\begin{align}
    \mathcal{R}(n_c)\approx
    \begin{cases}
    \delta N(n_{*1}, \varphi^a(n_k))
    +\delta N(n_{*3}, \varphi^a(n_{*2})) & \text{  for  }k < k_1\,,
    \\
    \delta N(n_{*3}, \varphi^a(n_k)) & \text{  for  }k > k_1\,.
    \end{cases}
\end{align}
The first stage is effectively driven by the $\phi$ field, with $\chi$ staying at $\chi_0$, and the third stage is effectively driven by the $\chi$ field, with $\phi$ quickly settling down to $\phi_c$. Thus, we may approximate $\mathcal{R}$ as
\begin{align}
    \mathcal{R}\approx
    \begin{cases}
    \frac{\partial N}{\partial \phi}\Big\vert_{n_k} \delta\phi(n_k) + \frac{\partial N}{\partial \chi}\Big\vert_{n_{*2}} \delta\chi(n_{*2})
    & \text{  for  }k < k_1
    \,,\\
    \frac{\partial N}{\partial \chi}\Big\vert_{n_k} \delta\chi(n_k)
    & \text{  for  }k > k_1
    \,.
    \end{cases}
\end{align}
The curvature power spectrum is then given by
\begin{align}
    \mathcal{P}_\mathcal{R} (k) = \frac{k^3}{2\pi^2} \langle  \mathcal{R}_{\bf k}(n_c) \mathcal{R}_{\bf k}^\dagger(n_c) \rangle\,,
\end{align}
where $\mathcal{R}_{\bf k}$ is the Fourier transform of $\mathcal{R}$.

Let us first compute the first derivatives of $N$, namely
\begin{align}
    \frac{\partial N}{\partial \phi}\Bigg\vert_{n_{k < k_1}}
    \,,\qquad
    \frac{\partial N}{\partial \chi}\Bigg\vert_{n_{k > k_1}}
    \,,\qquad
    \frac{\partial N}{\partial \chi}\Bigg\vert_{n_{*2}}
    \,.
\end{align}
Recalling that, during the first stage,
\begin{align}
    \frac{\phi'}{M_{\rm P}} \approx 2\sqrt{\frac{2}{3}}\left(
    \frac{1}{1-F} + 3 q F
    \right)\,,
\end{align}
in the large-$\mu^2$ limit, we find
\begin{align}
    N=\int_{n_i}^{n} dn' \approx
    -\frac{3}{4}\int_{F_i}^{F} F^{-1}(F-1)\left[
    1-3 q F(1-F)
    \right]dF\,,
\end{align}
where we have used $dF=\sqrt{2/3}(F/M_{\rm P})d\phi$. We thus obtain
\begin{align}
    N(\text{during Stage 1}) &\approx \frac{3}{4}\left(F_i - F - \ln\frac{F_i}{F}\right)
    \nonumber\\&\qquad
    +\frac{3}{4}q \left(F_i-F\right)\left(
    3+F_i^2 - 3F_i + F_i F - 3F + F^2
    \right)\,.
\end{align}
Hence,
\begin{align}
    \frac{\partial N}{\partial \phi}\Bigg\vert_{n_{k < k_1}} \approx
    -\frac{3}{4}\sqrt{\frac{2}{3}}\frac{F_k-1}{M_{\rm P}}
    -q\frac{9}{4}\sqrt{\frac{2}{3}}\frac{F_k(F_k-1)^2}{M_{\rm P}}
    \,,
\end{align}
where $F_k$ is $F$ at $n=n_k$.
Alternatively, we could have started with $F'\approx 4F(1+3q F(1-F))/(3(1-F))$ which leads to the same result.
During the final, third stage, $\phi$ quickly settles down to $\phi_c$, while $\chi$ satisfies
\begin{align}
    \chi' \approx
    4\xi \left\{
    \chi_0 + \left[
    \frac{M_{\rm P}^2}{\xi\chi_g^2}\left(1-\frac{3q}{\mu^4}\right)-1
    \right]\chi
    \right\}\,,
\end{align}
from which we find
\begin{align}
    N(\text{during Stage 3}) &\approx
    \frac{1}{4\xi}\left[
    \frac{M_{\rm P}^2}{\xi\chi_g^2}\left(1-\frac{3q}{\mu^4}\right)-1
    \right]^{-1}
    \nonumber\\&\qquad\times
    \ln\left\{
    \frac{\chi_0+[M_{\rm P}^2(1-3q\mu^{-4})/(\xi\chi_g^2)-1]\chi}{M_{\rm P}^2(1-3q\mu^{-4})\chi_0/(\xi\chi_g^2)}
    \right\}\,.
\end{align}
We note again that the $q$ contribution is highly suppressed by $\mu^{-4}$. In the following, therefore, we neglect this tiny contribution.
From the expression of $N(\text{during Stage 3})$, we find that
\begin{align}
    \frac{\partial N}{\partial \chi} \approx
    \frac{1}{4\xi}\left[
    \chi_0 + \left(\frac{M_{\rm P}^2}{\xi\chi_g^2}-1\right)\chi
    \right]^{-1}
    \,.\label{eqn:dNchi}
\end{align}
Finally, recalling from Eq.~\eqref{eqn:chisolS3} with the $q\mu^{-4}$ term neglected, {\it i.e.},
\begin{align}
    \chi(n)\approx
    \frac{\chi_0}{1-\xi\chi_g^2/M_{\rm P}^2}\left\{
    \exp\left[
    \frac{4(1-\xi\chi_g^2/M_{\rm P}^2)}{\chi_g^2/M_{\rm P}^2}(n-n_{*2})
    \right] - \frac{\xi\chi_g^2}{M_{\rm P}^2}
    \right\}
    \,,\label{eqn:chiS3app}
\end{align}
we obtain, by inserting this solution into Eq.~\eqref{eqn:dNchi}, that
\begin{align}
    \frac{\partial N}{\partial \chi}\Bigg\vert_{n_{k > k_1}}
    &\approx
    \frac{1}{4\xi\chi_0}\left[
    1+\frac{M_{\rm P}^2}{\xi\chi_g^2}\left\{
    \exp\left[
    \frac{4(1-\xi\chi_g^2/M_{\rm P}^2)}{\chi_g^2/M_{\rm P}^2}(n-n_{*2})
    \right] - \frac{\xi\chi_g^2}{M_{\rm P}^2}
    \right\}
    \right]^{-1}
    \,,\\
    \frac{\partial N}{\partial \chi}\Bigg\vert_{n_{*2}}
    &\approx
    \frac{\chi_g^2}{4\chi_0 M_{\rm P}^2}
    \,.\label{eqn:NchiN2}
\end{align}

Next, let us evaluate $\langle \delta\phi_{\bf k} \delta\phi_{\bf k'}^\dagger \rangle$ and $\langle \delta\chi_{\bf k} \delta\chi_{\bf k'}^\dagger \rangle$, where $\delta\phi_{\bf k}$ and $\delta\chi_{\bf k}$ are the Fourier transforms of the field fluctuations $\delta\phi$ and $\delta\chi$, respectively. These two-point correlation functions, if evaluated at the horizon crossing, take the standard form (see, {\it e.g.}, Refs.~\cite{Sasaki:1995aw,Nakamura:1996da,Gong:2002cx}):
\begin{align}
    \langle \delta\phi_{\bf k}\delta\phi_{\bf k'}^\dagger \rangle &=
    \delta({\bf k}-{\bf k'})G^{\phi\phi}\frac{H^2}{2k^3}\Bigg\vert_{n_k}
    \,,\\
    \langle \delta\chi_{\bf k}\delta\chi_{\bf k'}^\dagger \rangle &=
    \delta({\bf k}-{\bf k'})G^{\chi\chi}\frac{H^2}{2k^3}\Bigg\vert_{n_k}
    \,,
\end{align}
where $G^{ab}$ is the inverse of the field-space metric; in our case, $G^{\phi\phi} = 1$ and $G^{\chi\chi} = F$.
There is still one more two-point function we need in order to compute the curvature power spectrum, which is $\langle \delta\chi_{\bf k} \delta\chi_{\bf k'}^\dagger \rangle$ at the end of the second stage, or, equivalently, at the start of the third stage. As the mode is in the superhorizon region, this quantity is not given by the standard form of $H^2/(2k^3)$ at $n_{*2}$. To obtain $\langle \delta\chi_{\bf k} \delta\chi_{\bf k'}^\dagger \rangle$ at $n_{*2}$, let us note that, under our assumption,
\begin{align}
    \langle \delta\chi_{\bf k} \delta\chi_{\bf k'}^\dagger \rangle\big\vert_{n_{*2}}
    \approx
    \langle \delta\chi_{\bf k} \delta\chi_{\bf k'}^\dagger \rangle\big\vert_{n_{*1}}
    \,.
\end{align}
Perturbing the background equation of motion for the $\chi$ field during the first stage gives
\begin{align}
    3H\dot{\delta \chi} \approx -FV_{{\rm E},\chi\chi}\delta\chi\,.
\end{align}
Since
\begin{align}
    V_{{\rm E},\chi\chi} \approx
    3\xi M^2 \left(\frac{1}{F}-9q\right)
    \,,\quad
    H^2 \approx
    \frac{M^2}{4}\left(1-\frac{2}{F}-6q F\right)\,,
\end{align}
during the first stage, we get
\begin{align}
    \delta\chi_{*1} \approx
    \delta\chi e^{-4\xi\Xi}\,,
\end{align}
where
\begin{align}
    \Xi &=
    n_{*1} - n - \frac{3}{2}\ln\left[
    \frac{6-3F_{*1}}{6-3F_{*1}+4(n-n_{*1})}
    \right]
    \nonumber\\&\quad\quad
    +\frac{q}{6}\bigg\{
    \frac{27(F_{*1}^3-32)}{3F_{*1}-6-4(n-n_{*1})}
    -\frac{9(F_{*1}^3-32)}{F_{*1}-2}
    -4(n^2-n_{*1}^2)
    \nonumber\\&\quad\quad\quad
    +2n(3F_{*1}-42+4n_{*1})
    +n_{*1}(84-6F_{*1}-8n_{*1})
    \nonumber\\&\quad\quad\quad
    -270\ln\left[\frac{6-3F_{*1}}{6-3F_{*1}+4(n-n_{*1})}\right]
    \bigg\}
    \,.
\end{align}

There is another way to compute $\langle\delta\chi^2\rangle$, which is adopted in Ref.~\cite{Wang:2024vfv}. Perturbing the equation of motion for the $\chi$ field, without the slow-roll approximation, gives
\begin{align}
    \ddot{\delta\chi} + \left(3H-\frac{\dot{F}}{F}\right)\dot{\delta\chi}+FV_{{\rm E},\chi\chi}\delta\chi = 0\,.
\end{align}
In terms of the number of $e$-folds, the equation can be expressed as
\begin{align}
    \delta\chi'' + \left[
    3-\frac{(F/H)'}{F/H}
    \right]\delta\chi'
    +\frac{FV_{{\rm E},\chi\chi}}{H^2}\delta\chi = 0\,.
\end{align}
Defining $X \equiv \delta \chi / Y$, where $Y \equiv Y_0\sqrt{f/f_0}e^{-3(n-n_0)/2}$ with $f\equiv F/H$, $n_0$ being some reference number of $e$-folds, and the $f_0$ and $Y_0$ being the values of $f$ and $Y$ at $n=n_0$, the equation can be re-written as follows:
\begin{align}
    X'' + z^2 X = 0\,,
\end{align}
where
\begin{align}
    z^2 \equiv -\frac{9}{4} + 2\frac{f'}{f} - \frac{3}{4}\left(\frac{f'}{f}\right)^2 + \frac{FV_{{\rm E},\chi\chi}}{H^2}\,.
\end{align}
Note that $FV_{{\rm E},\chi\chi} \approx 3M^2\xi(1-9q F)$.
If we assume that $z$ slowly varies with time, {\it i.e.}, if we neglect the time variation of $f'/f$ and $FV_{{\rm E},\chi\chi}/H^2$, we can approximate the solution as
\begin{align}
    \delta\chi &\approx
    \delta\chi_0 \sqrt{\frac{f}{f_0}}e^{-3(n-n_0)/2} \cos\left[
    z(n-n_0)
    \right]
    \nonumber\\&\quad
    +\sqrt{\frac{f}{f_0}}e^{-3(n-n_0)/2}\frac{1}{z_0}\left\{
    \delta\chi_0' + \delta\chi_0\left[
    \frac{3}{2}-\left(\frac{f'}{2f}\right)_0
    \right]
    \right\}\sin\left[
    z(n-n_0)
    \right]\,,
\end{align}
where $\delta\chi_0'\equiv (d\delta\chi/dn)\vert_{n=n_0}$.
Thus, squaring it and taking the average, we obtain
\begin{align}
    \langle \delta\chi^2 \rangle \bigg\vert_{*1} \approx
    \langle \delta\chi^2 \rangle \bigg\vert_{n_k}
    \frac{(F/H)_{*1}}{(F/H)_{n_k}}e^{-3(n_{*1}-n_k)}\,.
\end{align}
We again stress that, to arrive at this result, we have assumed that the change of $z$ in time is negligible. In other words, we have neglected the time variation of $F/H$ and $FV_{{\rm E},\chi\chi}/H^2$; however, the absolute difference between the value of $F/H$ at $n_{*1}$ and the value of $F/H$ at $n_k$ has been maintained.
The former approach does not capture the non-slow-roll phase near the end of the first stage, but it correctly includes the evolution of $F$, $H$, and $V_{{\rm E},\chi\chi}$. On the other hand, the latter approach misses the time evolution of $F$, $H$, and $V_{{\rm E},\chi\chi}$ although it maintains the value difference in them between the end-of-the-first-stage point and the horizon-crossing point, while it takes into account the acceleration term in $\delta\chi$. The correct numerical results would sit somewhere between these two approaches.

We can now compute the power spectrum of primordial curvature perturbations:
\begin{align}
    \mathcal{P}_\mathcal{R} (k < k_1) &\approx
    \left(\frac{H(n_k)}{2\pi}\right)^2\left[
    \left(\frac{\partial N}{\partial \phi}\Bigg\vert_{n_k}\right)^2
    +F(n_k)\Psi\left(\frac{\partial N}{\partial \chi}\Bigg\vert_{n_{*2}}\right)^2
    \right]
    \,,\\
    \mathcal{P}_\mathcal{R} (k > k_1) &\approx
    \left(\frac{H(n_k)}{2\pi}\right)^2
    F(n_k)\left(\frac{\partial N}{\partial \chi}\Bigg\vert_{n_k}\right)^2
    \,,
\end{align}
where $\Psi$ is the factor related to $\langle \delta\chi^2 \rangle$. We discussed two different treatments, and for the first one, it is given by
\begin{align}
    \Psi = e^{-8\xi\Xi}\,,
\end{align}
while for the second treatment, it is given by
\begin{align}
    \Psi = \frac{(F/H)_{*1}}{(F/H)_{n_k}}e^{-3(n_{*1}-n_k)}\,.
\end{align}
In the expressions for the power spectrum, there appear $n_{*1}$ and $n_{*2}$. One may replace them by $n$, $k_1/k$, and $\Delta n(\text{Stage 2})$ as 
\begin{align}
    n_{*1} = n + \ln\frac{k_1}{k}\,,
\end{align}
and
\begin{align}
    n_{*2} = n + \ln\frac{k_1}{k} + \Delta n(\text{Stage 2})
    \,.
\end{align}
Then, the curvature power spectrum is given in terms of $\ln(k_1/k)$ instead of $n$.
Note also that
\begin{align}
    \frac{k_1}{k_p} = \mu^{-2}\exp\left(
    \Delta n_{\rm CMB} - \Delta n(\text{Stage 3})
    \right)\,,
\end{align}
where $k_p$ is the pivot value of comoving wavenumber to fit the CMB observation, and $\Delta n_{\rm CMB}$ is the total number of $e$-folds for $k_p$ from the horizon crossing to the end of inflation.
We remind that
\begin{align}
    \Delta n(\text{Stage 2}) &\approx
    \frac{1}{3}\ln \mu^2
    \,,\\
    \Delta n (\text{Stage 3}) &\approx
    \frac{1}{4}\left(\frac{\chi_g}{M_{\rm P}}\right)^2
    \left(1-\xi\frac{\chi_g^2}{M_{\rm P}^2}\right)^{-1}
    \ln\left[
    \frac{1}{2\sqrt{2}}\frac{\chi_g^2}{M_{\rm P}\chi_0}
    \right]\,.
\end{align}

We have, in total, six model parameters, namely $m$, $\lambda$, $M$, $\chi_0$, $\xi$, and $q$. For a given value of $\Delta n_{\rm CMB}$, however, one of these parameters, which is chosen to be $M$ in our analysis, gets fixed due to the Planck normalisation, {\it i.e.}, $\mathcal{P}_\mathcal{R} (k_p) = 2.1 \times 10^{-9}$. Furthermore, we introduce a convenient parameter $A\equiv \lambda M_{\rm P}^2/(\xi m^2)$ to replace $\lambda$. Thus, the free input parameters of our model become as follows:
\begin{align}
    \xi\,,\quad
    m\,,\quad
    \chi_0\,,\quad
    A\,,\quad
    q\,.
\end{align}
For convenience, here we again list the other quantities appearing in the expression of the power spectrum in terms of those free parameters:
\begin{align}
    \chi_g \equiv \frac{m}{\sqrt{\lambda}}
    \,,\quad
    V_m = \frac{m^4}{4\lambda} - \frac{1}{2}m^2\chi_0^2 + \frac{1}{4}\lambda\chi_0^4
    \,,\quad
    \mu = \sqrt{\frac{3M^2M_{\rm P}^2}{4V_m}}
    \,.
\end{align}

We are now in a position to examine the curvature power spectrum at the pivot scale, $\mathcal{P}_\mathcal{R}(k_p)$. Assuming that $k_p$ exits the horizon way before the end of the first stage, we may take $k_p \ll k_1$ and find
\begin{align}
    \mathcal{P}_\mathcal{R}(k_p \ll k_1) &\approx
    \frac{M^2}{384\pi^2M_{\rm P}^2}\left(
    -6+3F_{*1}-4\ln\frac{k_p}{k_1}
    \right)\left(
    3-3F_{*1}+4\ln\frac{k_p}{k_1}
    \right)^2\left(
    3F_{*1}-4\ln\frac{k_p}{k_1}
    \right)^{-1}
    \nonumber\\&\quad
    +q\frac{M^2}{1728\pi^2M_{\rm P}^2}\left(
    3F_{*1}-4\ln\frac{k_p}{k_1}
    \right)^{-2}\left(
    -3+3F_{*1}-4\ln\frac{k_p}{k_1}
    \right)
    \nonumber\\&\quad\quad\times
    \bigg\{
    729F_{*1}^2(F_{*1}-1)[2+F_{*1}(F_{*1}-4)]
    \nonumber\\&\quad\quad\quad
    -972F_{*1}(F_{*1}-1)[4+F_{*1}(4F_{*1}-15)]\ln\frac{k_p}{k_1}
    \nonumber\\&\quad\quad\quad
    +1296[-2+F_{*1}(19+7F_{*1}(F_{*1}-4))]\left(\ln\frac{k_p}{k_1}\right)^2
    \nonumber\\&\quad\quad\quad
    -576[19+4F_{*1}(5F_{*1}-14)]\left(\ln\frac{k_p}{k_1}\right)^3
    \nonumber\\&\quad\quad\quad
    +1536(5F_{*1}-7)\left(\ln\frac{k_p}{k_1}\right)^4
    -2048\left(\ln\frac{k_p}{k_1}\right)^5
    \bigg\}
    \,.\label{eqn:PRlong}
\end{align}
In the limit of $|\ln(k_p/k_1)| \gg 1$, we may approximate Eq.~\eqref{eqn:PRlong} as
\begin{align}
    \mathcal{P}_\mathcal{R} (k_p) \approx
    \frac{M^2}{24\pi^2 M_{\rm P}^2}\left(
    \ln\frac{k_p}{k_1}
    \right)^2
    +q\frac{8M^2}{27\pi^2 M_{\rm P}^2}\left(
    \ln\frac{k_p}{k_1}
    \right)^4\,.
\end{align}
The scalar spectral index $n_s$ at the pivot scale $k_p$ can be obtained from the definition
\begin{align}
    n_s - 1 \equiv \frac{d\ln\mathcal{P}_\mathcal{R}}{d\ln k}\Bigg\vert_{k=k_p}\,.
\end{align}
Up to the leading-order terms in the limit $|\ln(k_p/k_1)| \gg 1$, we find
\begin{align}
    n_s \approx 1
    +\frac{M^2}{12\pi^2 M_{\rm P}^2\mathcal{P}_\mathcal{R}(k_p)}\left(
    \ln\frac{k_p}{k_1}
    \right)\left[1+
    \frac{128q}{9}\left(
    \ln\frac{k_p}{k_1}
    \right)^2\right]
    \,.\label{eqn:nsapp}
\end{align}
It is evident from Eq.~\eqref{eqn:nsapp} that the presence of the $R^3$ term lowers (raises) the value of $n_s$ for $q > 0$ ($q < 0$). This result is consistent with the rough estimation shown in Refs.~\cite{Pi:2017gih,Wang:2024vfv}.

We note that an alternative way to compute the spectral index is through the slow-roll parameters such that
\begin{align}
    n_s \approx 1 - 6\epsilon_V + 2\eta_V\,,
\end{align}
where, as the scale of interest exists the horizon during the first stage of inflation, the potential slow-roll parameters are given by
\begin{align}
    \epsilon_V \equiv \frac{M_{\rm P}^2}{2}\left(
    \frac{V_{{\rm E},\phi}}{V_{\rm E}}
    \right)^2
    \,,\qquad
    \eta_V \equiv M_{\rm P}^2\frac{V_{{\rm E},\phi\phi}}{V_{\rm E}}
    \,.
\end{align}
We then obtain
\begin{align}
    n_s &\approx
    \frac{1}{3[1+(F_p-1)^2\mu^2]^2}\left[
    -5+2(11F_p^2-2F_p-5)\mu^2+(F_p-1)^2(3F_p^2-14F_p-5)\mu^4
    \right]
    \nonumber\\&\quad
    -q\frac{8F_p(F_p-1)\mu^6}{[1+(F_p-1)^2\mu^2]^3}\left[
    (F_p-3)(F_p-1)^4 + 3(5F_p-3)\mu^{-4} + 4(F_p-3)(F_p-1)^2\mu^{-2}
    \right]\,,
    \label{eqn:nsapp2}
\end{align}
where $F_p$ is the value of $F$ at the pivot scale $k_p$.
Similarly, one may compute the tensor-to-scalar ratio, which is given by
\begin{align}
    r &\approx 16\epsilon_V
    \nonumber\\ &\approx 
    \frac{64[(F_p-1)\mu^2-1]^2}{3[(F_p-1)^2\mu^2+1]^2}
    -q\frac{128F_p(F_p-1)^2\mu^6}{[(F_p-1)^2\mu^2+1]^3}
    \left[
    (F_p-1)^3
    -3\mu^{-4}
    -(F_p^2-5F_p+4)\mu^{-2}
    \right]
    \,.
    \label{eqn:r}
\end{align}
It is clear that the corrections of the Ricci cubic term to the spectral index and the tensor-to-scalar ratio are
\begin{align}
    \Delta n_s \approx -8qF_p\left(
    \frac{F_p-3}{F_p-1}
    \right)
    \,,\qquad
    \Delta r \approx -128q\left(
    \frac{F_p}{F_p-1}
    \right)
    \,,
\end{align}
where we have neglected terms that are suppressed by large $\mu^2$. Recalling that $F_p$ is the value of $F\equiv \exp(\sqrt{2/3}\phi/M_{\rm P})$ at the pivot scale where $\phi$ is larger than $M_{\rm P}$, we observe that $\Delta n_s \simeq -8qF_p$ and $\Delta r \simeq -128q$. Both the spectral index and the tensor-to-scalar ratio thus increase (decrease) for negative (positive) $q$ values, while the change in the tensor-to-scalar ratio is much smaller.

In line with our findings in the previous subsection, the analytical analysis of the cosmological perturbation presented in this subsection explicitly shows that the inclusion of the $R^3$ term has a noticeable effect on the CMB scale. In particular, the $R^3$ correction affects the curvature power spectrum; a negative (positive) $R^3$ term with $q < 0$ ($q > 0$) flattens (steepens) the curvature power spectrum on the CMB scale. It in turn has an effect of modifying the spectral index $n_s$; a positive (negative) $q$ lowers (raises) the spectral index value, compared to the Higgs-$R^2$ model case. Therefore, one may achieve a scenario where the produced PBHs constitute the whole dark matter abundance without being in tension with the observational bound on $n_s$ and $r$. In order to obtain quantitative values of $n_s$ and $r$, we need the value $F_p$ which is to be substituted into Eqs.~\eqref{eqn:nsapp2} and \eqref{eqn:r}. To find $F_p$ or, equivalently, the $\phi$-field value at the pivot scale for a fixed parameter set, we utilise the slow-roll-approximated expression for the normalisation $\mathcal{P}_{\mathcal{R}} \approx V_{\rm E}/(24\pi^2 M_{\rm P}^4 \epsilon_V) \approx 2.1 \times 10^{-9}$.
One may alternatively choose a specific value for the total number of $e$-folds between the pivot scale and the end of inflation\footnote{
Considering a concrete reheating physics would determine the number of $e$-folds. However, we remain agnostic to such a specification, allowing variation in the total number of $e$-folds during inflation within the range $50 \lesssim N \lesssim 60$.
} and then use the parameter $M$, which controls the characteristic inflationary energy scale during Stage 1, to match the normalisation $\mathcal{P}_{\mathcal{R}} \approx 2.1 \times 10^{-9}$. We summarise the values of $n_s$ and $r$ obtained from the analytical expressions \eqref{eqn:nsapp2} and \eqref{eqn:r} in Table~\ref{tab:nsr} for four benchmark cases outlined in Table~\ref{tab:parameters}.
In the next subsection, we verify our analytical understanding by numerically solving the system.

\subsection{Numerical Treatments}
\label{subsec:enhancement}
Having the analytical understanding of the effect of the Ricci cubic term, in particular, on the curvature power spectrum $\mathcal{P}_{\mathcal{R}}$, the spectral index $n_s$, and the tensor-to-scalar ratio $r$, we now present results of the numerical calculation without approximations. We closely follow Ref.~\cite{Wang:2024vfv}. All the relevant equations to be solved are summarised in Appendix A of Ref.~\cite{Wang:2024vfv}. For completeness, we briefly explain the numerical method. For a chosen parameter set, we first solve the background equations \eqref{eqn:bkgeomH}, \eqref{eqn:bkgeomphi} and \eqref{eqn:bkgeomchi} until the end of inflation determined by the condition $\epsilon = 1$ at the end of the third stage of inflation. For the perturbations, we solve the linear perturbation equations in Fourier space on the flat slicing \cite{Wang:2024vfv,Sasaki:1995aw},
\begin{align}
    D_t^2\delta\varphi_k^a + 3H D_t\delta\varphi_k^a + \frac{k^2}{a^2}{\delta\varphi_k^a}+V_{\rm E}{}^{;a}_{\ \ ;b}\delta\varphi_k^b - R^a{}_{bcd}\dot\varphi^b\dot\varphi^c\delta\varphi_k^d-\frac{1}{a^3}D_t\left(\frac{a^3}{H}\dot\varphi^a\dot\varphi^b\right)h_{bc}\delta\varphi^b = 0
    \,,
\end{align}
where $\varphi^a=\{\phi,\chi\}$ are the background fields, $\delta\varphi^a_k=\{\delta\phi_k,\delta\chi_k\}$ are the field perturbations, $D_t$ is the covariant derivative in the field space, {\it e.g.}, $D_t\varphi^a = \dot{\varphi}^a + \Gamma^a_{bc}\varphi^b\varphi^c$, and $\Gamma^a_{bc}$ and $R^a{}_{bcd}$ are, respectively, the Christoffel symbols and the Riemann tensor associated with the field space metric $h_{ab}$. The initial conditions are taken to be the standard Bunch-Davies vacuum state in deep subhorizon; we choose 7 $e$-folds before the horizon crossing. The curvature perturbation is then given by $\mathcal{R} = (\phi'\delta\phi+F^{-1}\chi'\delta\chi)/(2\epsilon M_{\rm P}^2)$; recall that the prime denotes the derivative with respect to the number of $e$-folds. We evaluate the curvature perturbation at the end of inflation. Once the curvature power spectrum is obtained, we compute the scalar spectral index $n_s$ by taking the slope at the pivot scale of $k_{p} = 0.05 \, {\rm Mpc}^{-1}$,
\begin{align}
    n_s=1+\left.\frac{d\ln \mathcal{P}_{\mathcal{R}}(k)}{d\ln k}\right|_{k_{p}}
    \,.
\end{align}
For the tensor-to-scalar ratio $r$, we utilise the single-field approximation, $r = 16\epsilon$, which is reasonable in our case as, when the large-scale modes exit the horizon during the early time of the first stage of inflation, the dynamics of the background converges to the single-field slow-roll inflation.

\begin{figure}[t!]
	\centering
	\includegraphics[width=0.7\textwidth]{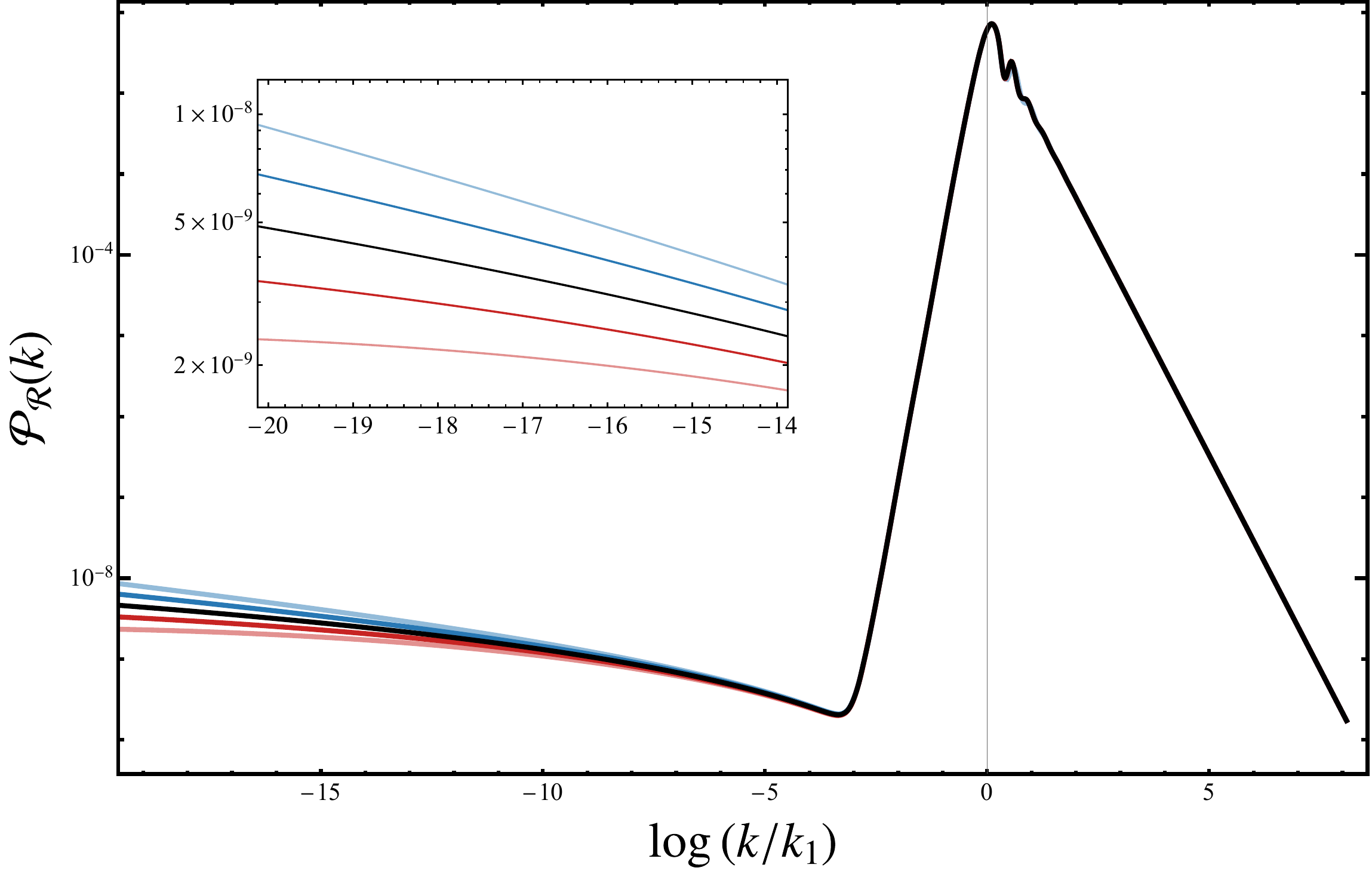}
	\caption{The curvature power spectrum $\mathcal{P}_\mathcal{R}$ is numerically obtained for five different $q$ values. The $q=0$ case (black) corresponds to the standard Higgs-$R^2$ model. The light (dark) red line corresponds to the $q = -4\times 10^{-5}$ ($q = -2\times 10^{-5}$) case, while the light (dark) blue line corresponds to the $q = 4 \times 10^{-5}$ ($q = 2 \times 10^{-5}$) case. The rest of the parameters are chosen as $\{M, m, \xi, A, \chi_0\} = \{2.2\times 10^{-5}M_{\rm P}, 6.6\times 10^{-6}M_{\rm P}, 0.3125, 1.5, 0.062M\}$ and remain unchanged to show the effect of the $R^3$ term.}
	\label{fig:PRq}
\end{figure}
\begin{figure}[ht!]
	\centering
	\includegraphics[width=0.75\textwidth]{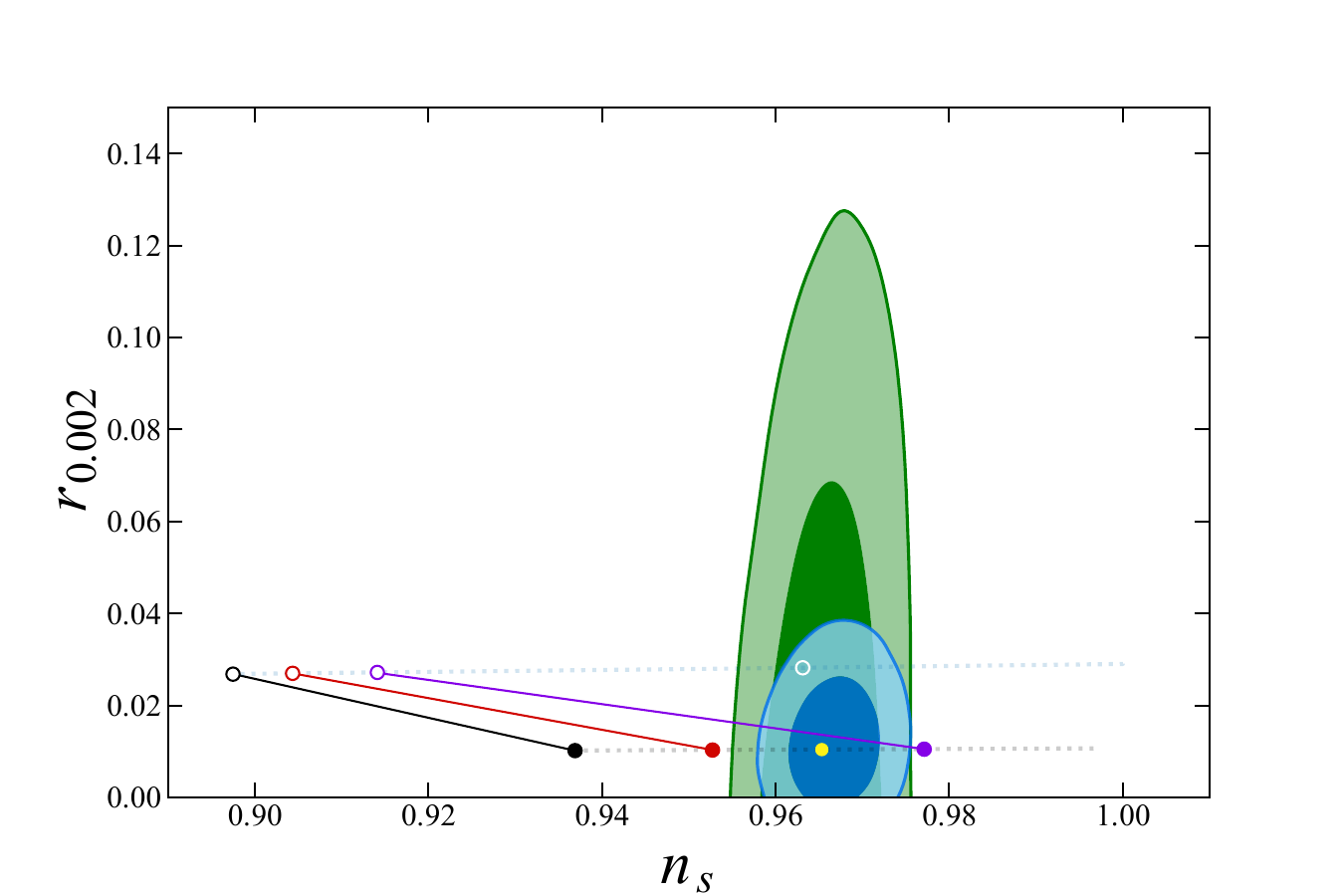}
	\caption{Values of the spectral index $n_s$ and the tensor-to-scalar ratio $r$ are shown together with the Planck constraint~\cite{Planck:2018vyg} (green contours) and the improved constraints of Planck and BICEP/Keck~\cite{BICEP:2021xfz} (blue contours). The tensor-to-scalar ratio (spectral index) is evaluated at the pivot scale of $k_p = 0.002\,{\rm Mpc}^{-1}$ ($k_p = 0.05\,{\rm Mpc}^{-1}$). The lower solid dots are for the parameter set explored in Fig.~\ref{fig:PRq} with $q = 0$ (black), $q = -2 \times 10^{-5}$ (red), and $q = -4 \times 10^{-5}$ (purple). The yellow solid dot represents the value of $q = -3.2 \times 10^{-5}$.  The upper hollow dots are the results for Case 2b presented in Table~\ref{tab:parameters} with $q = 0$ (black), $q = -2\times 10^{-5}$ (red), and $q = -4\times 10^{-5}$ (purple). The white hollow dot represents the value of $q = -1.1 \times 10^{-4}$. The dotted lines represent the behaviour of $n_s$ and $r$ when $q$ negatively increases.}
	\label{fig:nsrq}
\end{figure}

In Fig.~\ref{fig:PRq}, the numerically obtained curvature power spectrum $\mathcal{P}_{\mathcal{R}}$ is shown for five different values of $q$, while the rest of the parameters are fixed as $\{M, m, \xi, A, \chi_0\} = \{2.2\times 10^{-5}M_{\rm P}, 6.6\times 10^{-6}M_{\rm P}, 0.3125, 1.5, 0.062M\}$. The black curve corresponds to the case without $R^3$ term ($q = 0$), thus reduces to the standard Higgs-$R^2$ case. The light red (blue) curve corresponds to $q = -4 \times 10^{-5}$ ($q = 4 \times 10^{-5}$), while the dark red (blue) curve is for $q = -2 \times 10^{-5}$ ($q = 2 \times 10^{-5}$).
It is evident that the slope of the curvature power spectrum near the CMB scale becomes milder for a negative $R^3$ correction.
In contrast, small-scale behaviour is not greatly affected by the presence of the $R^3$ term. In other words, the curvature enhancement remains intact as in the standard Higgs-$R^2$ case~\cite{Wang:2024vfv}.

One of the most interesting features of our model is that the $R^3$ correction term could shift the $n_s$--$r$ towards the favoured region of the joint constraints from Planck and BICEP/Keck~\cite{Planck:2018vyg,BICEP:2021xfz}. For the same parameter set used in Fig.~\ref{fig:PRq}, the values of $n_s$ and $r$ are shown in Fig.~\ref{fig:nsrq} for $q = 0$ (black solid dot), $q = -2 \times 10^{-5}$ (red solid dot), and $q = -4 \times 10^{-5}$ (purple solid dot). As expected, the $R^3$ term with $q < 0$ raises the value of $n_s$; when $q = -3.2 \times 10^{-5}$ (yellow solid dot), we see that the spectral index becomes compatible with the observation. The upper hollow dots in Fig.~\ref{fig:nsrq} represent Case 2b in Table~\ref{tab:parameters} with $q = 0$ (black), $q = -2 \times 10^{-5}$ (red), and $q = -4 \times 10^{-5}$ (purple). In this case, the coefficient of the $R^3$ term, $q$, needs to be negatively larger in order to make $n_s$ compatible with the observational bound; when $q = -1.1 \times 10^{-4}$, we have $n_s \approx 0.963$ (see Table~\ref{tab:nsr}).\footnote{
We note that the $n_s$ value preferred by the recent ACT analysis \cite{ACT:2025fju,ACT:2025tim}, namely $n_s = 0.974 \pm 0.003$, can also be made compatible with the inclusion of the $R^3$ term; see, {\it e.g.}, the purple line in Fig.~\ref{fig:nsrq}.
}
The change in the tensor-to-scalar ratio $r$ is too small to be visible which also agrees with our expectation.
As presented in Table~\ref{tab:nsr}, the numerical analysis aligns well with our analytical understanding presented in the previous subsection.

\section{PBHs and GWs}
\label{sec:PBHandGW}
When the enhanced curvature perturbations re-enter the horizon, PBHs will form from the gravitational collapse of the large overdense region. For definiteness, we consider the case where the modes with enhanced curvature perturbations re-enter the horizon during the radiation-dominated era. The PBH mass is proportional to the horizon mass $M_H$ such that~\cite{Carr:1975qj}
\begin{align}
    M_{\rm PBH} = \gamma M_H
    \simeq
    \gamma M_\odot \left(
    \frac{g_*(T)}{10.75}
    \right)
    \left(
    \frac{4.2\times10^6\,{\rm Mpc}^{-1}}{k}
    \right)^2
    \,,
\end{align}
where $\gamma \sim 0.2$, $M_\odot$ is the solar mass, and $g_*(T)$ is the effective relativistic degrees of freedom at temperature $T$ at the time of horizon re-entry.
The abundance of PBHs can be estimated by adopting the Press-Schechter formalism.
Taking the Gaussian distribution of density perturbations, the PBH abundance at the time of formation can be estimated as
\begin{align}
    \beta(M_{\rm PBH}) = 
    \frac{\gamma}{\sqrt{2\pi\sigma_\delta^2}}
    \int_{\delta_{\mathrm{th}}}^{\infty}
    \exp\left(
    -\frac{\delta^2}{2\sigma_{\delta}^2}
    \right) 
    d\delta
    =
    \gamma~\mathrm{erfc}\left(
    \frac{\delta_{\mathrm{th}}}{\sqrt{2}\sigma_{\delta}}
    \right)
    \,,
\end{align}
where the variance of the density perturbation is given by
\begin{align}
    &\sigma_\delta^2 = 
    \int_0^\infty W^2(k; R) \mathcal{P}_\delta(t, k) d(\ln k)
    \,,\label{eqn:sig}\\
    &\mathcal{P}_{\delta}(k) = 
    \left(\frac{4}{9}\right)^2
    (kR)^4\mathcal{P}_\mathcal{R}(k)
    \,,
\end{align}
with $R = (aH)^{-1}$ being the comoving horizon size and $W^2(k; R)$ the window function; the Gaussian window function $W^2(k; R)=\exp\left(-k^2R^2/2\right)$ is adopted in this work.
The PBH abundance is usually presented with the quantity $f_{\rm PBH}$, which is the ratio of the PBH relic to the cold dark matter relic. Considering the evolution of the universe after the PBH formation, we can express $f_{\rm PBH}$ today as~\cite{Ando:2018qdb}
\begin{align}
    f_{\rm PBH}(M_{\rm PBH}) \equiv
    \frac{\Omega_{\rm PBH}}{\Omega_{\rm DM}} \simeq
    \left(
    \frac{\beta(M)}{1.6\times10^{-9}}
    \right)
    \left(
    \frac{10.75}{g_*(T)}
    \right)^{1/4}
    \left(
    \frac{0.12}{\Omega_{\rm DM}h^2}
    \right)
    \left(
    \frac{M_{\odot}}{M_{\rm PBH}}
    \right)^{1/2}
    \,.
\end{align}
\begin{table}[t!]
    \centering
    \begin{tabular}{c|c|c|c|c}
    \hline
    Case & 1a & 1b & 2a & 2b \\
    \hline
    $m/M_{\rm P}$ & $4.9\times10^{-6}$& $\ 6.6\times10^{-6}\ $ & $4\times10^{-6}$ & $1\times10^{-5}$\\
      
    $\xi$ & $0.077$ & $0.3125$ & $0.125$ & $0.3125$\\
     
    A & $2.8612$ & $1.5$ & $1.70455$ & $1.28$\\
     
    $\chi_0/M$ & $0.522725$ & $0.123865$ & $0.156709$ & $0.116219$\\

    $q$ & $-3.2\times 10^{-5}$& $-3.0\times 10^{-5}$& $-1.1\times 10^{-4}$& $-1.1\times 10^{-4}$\\
    \hline
    \end{tabular}%
    \caption{Input model parameters for the four benchmark cases used for the estimation of PBH abundance. The parameter $M$ is used to match the Planck normalisation, {\it i.e.}, $\mathcal{P}_\mathcal{R}(k_p) \approx 2.1 \times 10^{-9}$; the value of $M$ is given by $2.2 \times 10^{-5} M_{\rm P}$ ($3.62 \times 10^{-5} M_{\rm P}$) for Case 1a and Case 1b (Case 2a and Case 2b).}
    
    \label{tab:parameters}
\end{table}
\begin{table}[t!]
    \centering
    \begin{tabular}{c|c|c|c|c}
    \hline
    Case & 1a & 1b & 2a & 2b \\
    \hline
    $n_{s}$ (derived numerically)& $0.96536$& $0.96536$& $0.96321$& $0.96315$\\
     $n_{s}$ (derived from Eq.~\eqref{eqn:nsapp2}) & $0.96558$& $0.96558$& $0.96530$& $0.96524$\\
    $r$ (derived numerically)& $0.010404$& $0.010404$& $0.030888$& $0.028199$\\
     $r$ (derived from Eq.~\eqref{eqn:r})& $0.010803$& $0.010803$& $0.027509$& $0.027519$\\
    \hline
    \end{tabular}%
    \caption{Spectral index $n_s$ and tensor-to-scalar ratio $r$ for the four benchmark cases of Table~\ref{tab:parameters} at the pivot scale. Both the fully numerical results and the analytical results are presented. 
    The total numbers of $e$-folds are estimated to be 60.09 (Case 1a), 60.02 (Case 1b), 61.60 (Case 2a), and 60.90 (Case 2b), and the numbers of $e$-folds during Stage 1 are given by 36.77 (Case 1a), 36.73 (Case 1b), 24.02 (Case 2a), and 24.16 (Case 2b). The inclusion of the $R^3$ term makes $n_s$ and $r$ become compatible with the observational constraints, $0.958 \leq n_s \leq 0.975$ (95\% C.L.)~\cite{Planck:2018jri,BICEP:2021xfz} and $r \leq 0.036$ (95\% C.L.)~\cite{Planck:2018jri,BICEP:2021xfz}.}
    \label{tab:nsr}
\end{table}
Taking $\delta_{\rm th} = 0.3$ and $g_*$ according to Ref.~\cite{Franciolini:2022tfm}, we numerically estimate the PBH abundance for four benchmark cases outlined in Table~\ref{tab:parameters}.
As we discussed in the previous section, while the inclusion of the $R^3$ term has negligible effect on the small scales, its presence is crucial for the observationally-compatible spectral index $n_s$. Furthermore, the $R^3$ correction to the curvature power spectrum near the CMB scale demands adjustments of the other model parameters in order to match the Planck normalisation, {\it i.e.}, $\mathcal{P}_\mathcal{R} (k_p) \approx 2.1 \times 10^{-9}$. Table~\ref{tab:nsr} shows the values of the spectral index $n_s$ and the tensor-to-scalar ratio $r$ for the four benchmark cases. Due to the presence of the $R^3$ term, $n_s$ and $r$ may become compatible with the observational constraints, unlike in the standard Higgs-$R^2$ model \cite{Wang:2024vfv}.

\begin{figure}[t!]
    \centering
    \includegraphics[width=0.8\textwidth]{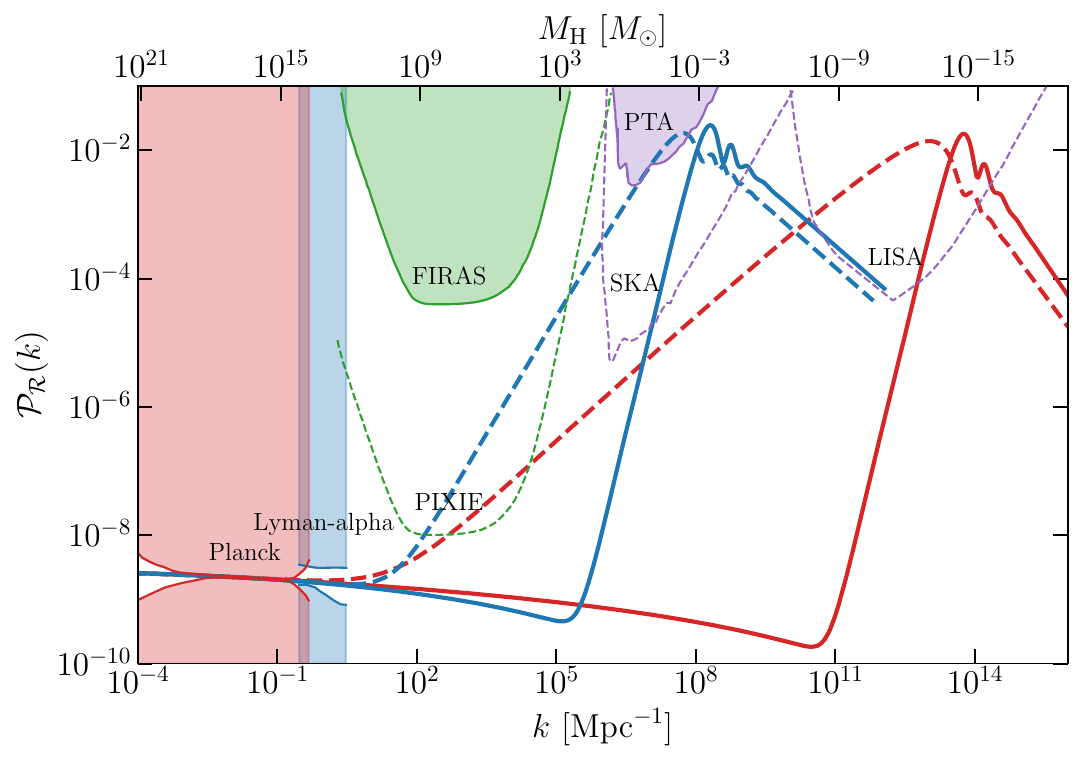}
    \caption{
 The curvature power spectrum is shown for the four benchmark cases outlined in Table~\ref{tab:parameters}. The dashed red (solid red) curve corresponds to Case 1a (Case 1b), while the dashed blue (solid blue) curve corresponds to Case 2a (Case 2b). Various constraints on the curvature power spectrum, which are adopted from Ref.~\cite{kavanagh2019}, are overlaid, including the current observational constraints of Planck satellite~\cite{Planck:2018jri}, Lyman-alpha forest~\cite{Bird:2010mp}, Far Infrared Absolute Spectrophotometer (FIRAS) experiment~\cite{Bianchini:2022dqh}, as well as the expected sensitivity of the future experiment Primordial Inflation Explorer (PIXIE)~\cite{Chluba:2013pya,Abitbol:2017vwa}. The purple-coloured region represents the range and magnitude of the curvature power spectrum for the results of Pulsar Timing Arrays (PTAs)~\cite{Iovino:2024tyg,NANOGrav:2023hvm,EPTA:2023xxk, Reardon:2023zen,Xu:2023wog}, while the purple dashed lines are the sensitivity curves of future measurements by Square Kilometre Array (SKA)~\cite{Weltman:2018zrl} and LISA~\cite{Baker:2019nia}.}
    \label{fig:PRplot}
\end{figure}
The numerical results of the curvature power spectrum for the four benchmark cases are presented in Fig.~\ref{fig:PRplot}. Case 1a and Case 1b are shown as the dashed red curve and the solid red curve, respectively, while Case 2a and Case 2b are respectively depicted by the dashed blue curve and the solid blue curve. Compared to the blue curves, the red curves have the peak of the curvature enhancement at a higher $k$ scale. The difference between the dashed and solid curves is the slope of the growth of the curvature perturbations. The scaling behaviour of the curvature power spectrum is mainly controlled by the $\xi$ parameter. Case 1b and Case 2b, for which $\xi=5/16$, feature the $k^3$ growth in the power spectrum, while Case 1a and Case 2a, which have $\xi < 3/16$, lead to a milder growth in the power spectrum. As a consequence, the dashed curves cross the PIXIE sensitivity line; we shall discuss this aspect in Sec.~\ref{sec:SD}.

\begin{figure}[t!]
    \centering
    \includegraphics[width=0.8\textwidth]{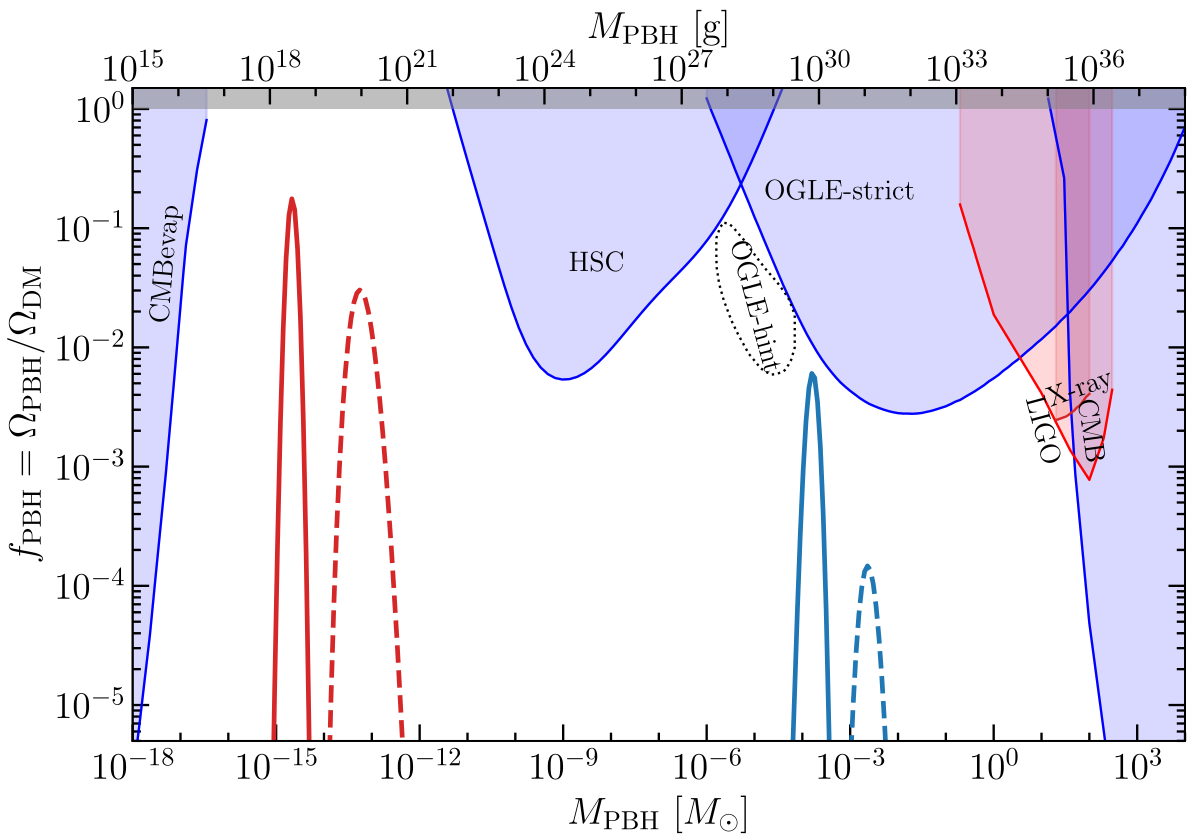}
    \caption{The PBH abundance $f_{\rm PBH}$ is computed from the curvature power spectrum for the four benchmark cases outlined in Table~\ref{tab:parameters}. The same colour scheme is taken as in Fig.~\ref{fig:PRplot}; the dashed red (solid red) curve corresponds to Case 1a (Case 1b), while the dashed blue (solid blue) curve corresponds to Case 2a (Case 2b). The colour-shaded regions depict various constraints adopted from Ref.~\cite{kavanagh2019}.}
    \label{fig:PBHplot}
\end{figure}
From the numerical results of the curvature power spectrum, we can compute the PBH abundance, $f_{\rm PBH}$, following the discussion above. The results for the four benchmark cases are presented in Fig.~\ref{fig:PBHplot}. As in Fig.~\ref{fig:PRplot}, the dashed (solid) red curve represents Case 1a (Case 1b), and the dashed (solid) blue curve depicts Case 2a (Case 2b).
Due to the broadness in the curvature power spectrum, the dashed curves (Case 1a and Case 2a) are always wider than the solid curves (Case 1b and Case 2b).
We stress again that, although the inclusion of the $R^3$ term does not heavily impact the small scales, and thus, the enhancement of curvature perturbations as well as the production of PBHs remain similar to the scenario of Higgs-$R^2$ inflation, the spectral index $n_s$ is now compatible with the latest observational data due to the presence of the $R^3$ term. Therefore, unlike the standard Higgs-$R^2$ model, the PBHs may be produced in the $10^{-16} \lesssim M_{\rm PBH}/M_\odot \lesssim 10^{-11}$ region where the total dark matter relic today could be accounted for by the produced PBHs, without being in tension with the $n_s$ bound.

The enhanced curvature perturbations may also source GWs~\cite{Matarrese:1997ay,Mollerach:2003nq,Ananda:2006af,Baumann:2007zm}. As the scalar mode induces the GWs at the second order in perturbation, they are often called scalar-induced, second-order GWs. Assuming again the radiation era at the time of GW generation, we may use the following semi-analytical expression for the GW density parameter~\cite{Kohri:2018awv}:
\begin{align}
    \Omega_{\rm GW,f} &= 
    \frac{1}{12}\int_0^\infty dv
    \int_{|1-v|}^{1+v} du
    \left(
    \frac{4v^2 - (1+v^2-u^2)^2}{4uv}
    \right)^2
    \mathcal{P}_\mathcal{R}(kv)\mathcal{P}_\mathcal{R}(ku)
    \nonumber\\&\quad\times
    \left(
    \frac{3(u^2+v^2-3)}{4u^3v^3}
    \right)^2
    \bigg[
    \left(
    -4uv+(u^2+v^2-3)\log\bigg\vert
    \frac{3-(u+v)^2}{3-(u-v)^2}
    \bigg\vert
    \right)^2
    \nonumber\\&\quad\quad
    +\pi^2(u^2+v^2-3)^2\theta(v+u-\sqrt{3})
    \bigg]\,,
\end{align}
where the subscript `f' denotes the fact that it is the quantity at the production time. The density parameter today is then given by
\begin{align}
    \Omega_{\rm GW} = \Omega_{\rm rad,0}\Omega_{\rm GW,f}\,,
\end{align}
where $\Omega_{\rm rad,0} \approx 0.9\times 10^{-4}$ is the current energy density parameter of radiation.

\begin{figure}[t!]
    \centering
    \includegraphics[width=0.8\textwidth]{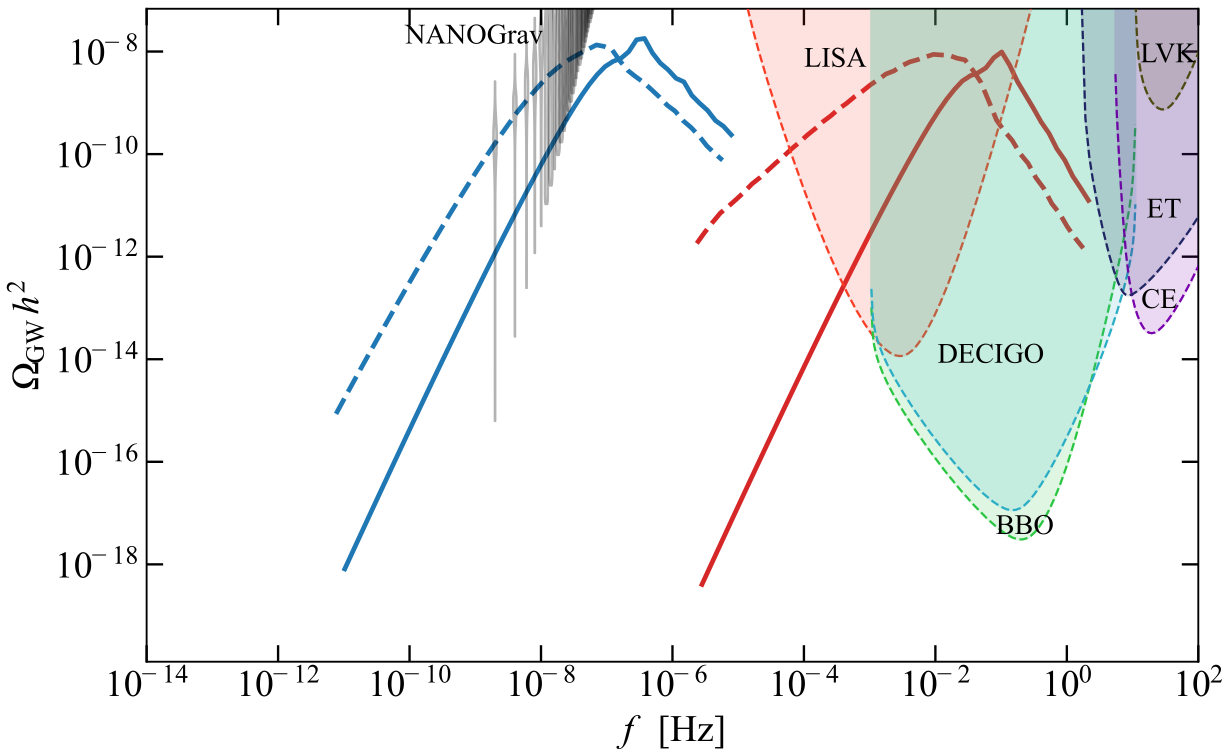}
    \caption{The spectrum of scalar-induced, second-order GWs, $\Omega_{\rm GW}h^2$, for the four benchmark cases outlined in Table~\ref{tab:parameters} is presented. The same colour scheme is taken as in Fig.~\ref{fig:PRplot}; the dashed red (solid red) curve corresponds to Case 1a (Case 1b), while the dashed blue (solid blue) curve corresponds to Case 2a (Case 2b). Case 1a and Case 1b fall into the nano-Hertz frequency range where the recently reported NANOGrav signal (gray violins) lies~\cite{NANOGrav:2023hvm}. On the other hand, Case 2a and Case 2b correspond to the deci-Hertz frequency range where future experiments including LISA~\cite{LISA:2017pwj}, DECIGO~\cite{Kawamura:2011zz}, and BBO~\cite{Crowder:2005nr} have the detectability. We also include the sensitivities of the LIGO-VIRGO-KAGRA (LVK) detector network~\cite{LIGOScientific:2014pky,VIRGO:2014yos,KAGRA:2018plz}, Einstein Telescope (ET)~\cite{Punturo:2010zz}, and Cosmic Explorer (CE)~\cite{Reitze:2019iox}.}
    \label{fig:GWplot}
\end{figure}
For the four benchmark cases presented in Table~\ref{tab:parameters}, the numerical results of the GW spectrum, $\Omega_{\rm GW}h^2$, are shown in Fig.~\ref{fig:GWplot}. Case 1a (dashed red) and Case 1b (solid red) have the GW spectrum in the deci-Hertz frequency range, while Case 2a (dashed blue) and Case 2b (solid blue) fall into the nano-Hertz frequency range, covering the stochastic GW signal recently reported by NANOGrav~\cite{NANOGrav:2023hvm}. Due to the broadness in the curvature power spectrum, Case 1b and Case 2b feature a sharper peak in the GW spectrum, compared to Case 1a and Case 2a.

\section{Spectral Distortions}
\label{sec:SD}
In this section, we comment on another potential observational signature of the enhanced curvature power spectrum, namely spectral distortions in CMB. Spectral distortions quantify deviations from the perfect black-body spectrum of the CMB photons. For a review, readers may refer to Refs.~\cite{Chluba:2019kpb,Lucca:2019rxf,Chluba:2019nxa} and references therein. 
Spectral distortions take various forms depending on the epoch and mechanism of their generation. The three most prominent types are the temperature shifts $g$-distortion, chemical potential $\mu$-distortion, and Compton $y$-distortion. In addition to these, there may also exist intermediate distortions, and such residual distortions are usually called the $r$-type distortions.
The total intensity of spectral distortions, $\mathcal{I}_{\rm tot}$, encompassing all these contributions, can then be expressed as
\begin{align}
    \mathcal{I}_{\rm tot}(z, x) = 
    \mathcal{I}_g(z, x) +
    \mathcal{I}_\mu(z, x) + 
    \mathcal{I}_y(z, x) + 
    \mathcal{I}_r(z, x)
    \,,
\end{align}
where $z$ denotes the redshift, and $x \equiv hf/(k_{\rm B}T_{\rm CMB})$ is the dimensionless frequency, with $h$, $k_{\rm B}$, $f$, and $T_{\rm CMB}$ being, respectively, the Planck constant, the Boltzmann constant, the frequency, and the present-day CMB temperature.

Amongst various sources for such deviations is the primordial curvature power spectrum from inflation.
Analytically, the spectral distortions for the CMB photon intensity spectrum can be obtained via the Green function method~\cite{Chluba:2013vsa,Lucca:2019rxf},
\begin{align}
    \mathcal{I}_{\rm tot}(z, x) =
    \int_{z}^\infty dz' G(z', x)\frac{dQ(z')/dz'}{\rho_\gamma(z')}
    \,,
\end{align}
where $\rho_\gamma$ is the photon energy density, $G$ is Green's function of the spectral distortion, and $dQ/dz$ is the heating rate, which is given by~\cite{Chluba:2012gq,Chluba:2013dna,Lucca:2019rxf,Fu:2020wkq}
\begin{align}
    \frac{dQ}{dz} = 
    4A^2\rho_\gamma
    \partial_z k_D^{-2}
    \int_{k_{\rm min}}^\infty dk \,
    \frac{k^4}{2\pi^2}
    \mathcal{P}_\mathcal{R}(k)
    e^{-2k^2/k_D^2}\,,
\end{align}
with $A$ being a normalisation factor, $k_D$ the photon damping scale, and $k_{\rm min} = 1\,{\rm Mpc}^{-1}$.
We see that a feature of the curvature power spectrum $\mathcal{P}_\mathcal{R}(k)$ would leave a distinctive imprint on the spectral distortions. Comparing the prediction of spectral distortions arising from the enhanced curvature power spectrum with those from the standard, simple power-law would provide a complementary observational prospect in addition to the scalar-induced GWs. As the next generation CMB experiments, such as PIXIE, are expected to probe the CMB spectrum for small distortions, understanding and forecasting spectral distortions from the enhanced curvature power spectrum would serve as a valuable asset. Recent similar studies include Refs.~\cite{Bae:2017tll,Byrnes:2018txb,Cabass:2018jgj,Zegeye:2021yml,Baur:2023naq,Mastache:2023cha,Tagliazucchi:2023dai}.

\begin{figure}[t!]
    \centering
    \includegraphics[width=0.8\textwidth]{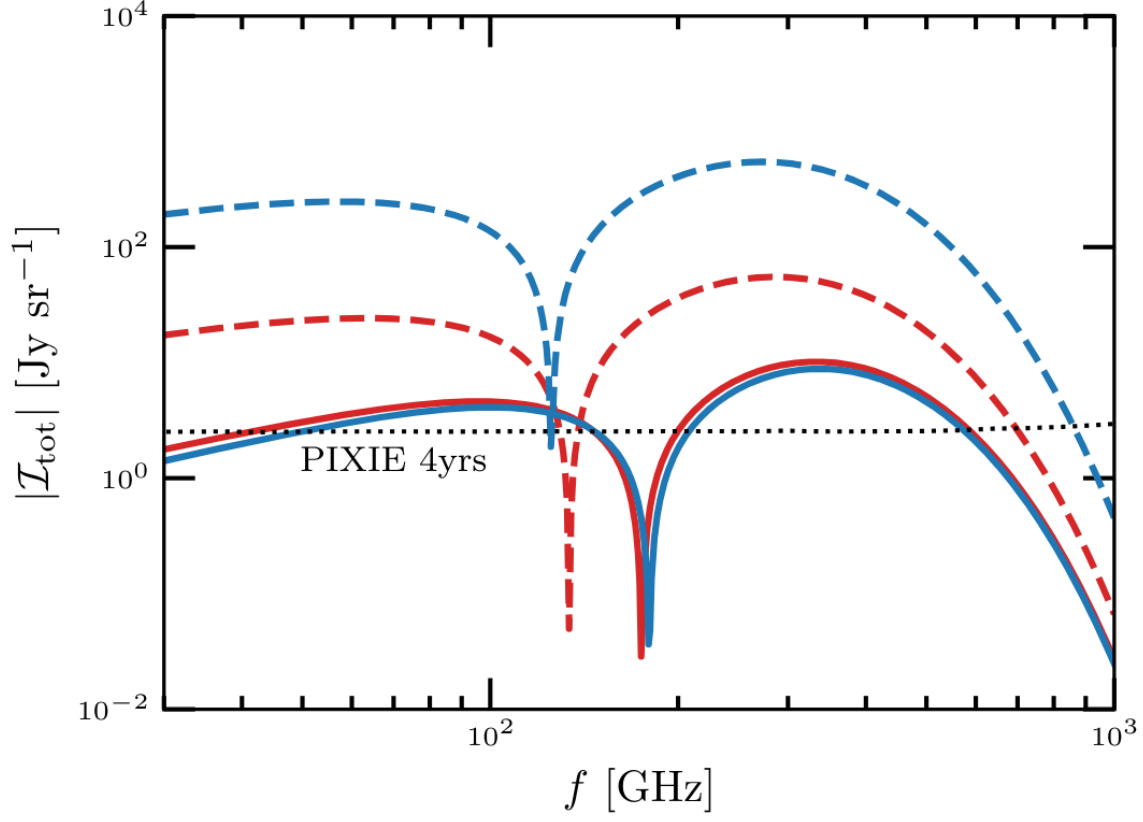}
    \caption{Absolute values of the spectral distortion intensity $\mathcal{I}_{\rm tot}$ are shown for the four benchmark cases given in Table~\ref{tab:parameters}. The same colour scheme is taken as in Fig.~\ref{fig:PRplot}; the dashed red (solid red) curve corresponds to Case 1a (Case 1b), while the dashed blue (solid blue) curve corresponds to Case 2a (Case 2b). The black dotted line represents the expected sensitivity of the future experiment PIXIE~\cite{Chluba:2019nxa,Abitbol:2017vwa}.}
    \label{fig:SDs}
\end{figure}
In this work, we utilise the publicly-available Boltzmann solver {\tt CLASS} \cite{Lesgourgues:2011re,Blas:2011rf,Lucca:2019rxf} to numerically compute the spectral distortions for the enhanced curvature power spectra obtained in Sec.~\ref{subsec:enhancement}. In Fig.~\ref{fig:SDs}, we present the absolute values of the total spectral distortion intensity for our four benchmark cases outlined in Table~\ref{tab:parameters}. The cases featuring a milder growth in the curvature power spectrum (Case 1a and Case 2a) exhibit larger spectral distortion intensities compared to those with the steeper growth (Case 1b and Case 2b). The expected sensitivity of PIXIE is also shown by the black dotted line in Fig.~\ref{fig:SDs}.
The difference in spectral distortion intensities between the cases with different values of $\xi$ can be understood from the scaling of the curvature power spectrum. For $\xi = 5/16$ (Case 1b and Case 2b), the sharp $k^3$ growth results in a narrower enhancement at much smaller scales, leading to smaller spectral distortions. On the other hand, for $\xi < 3/16$ (Case 1a and Case 2a), the broader enhancement shows the growth already near the CMB scale, generating larger distortions.

The detectability of spectral distortions provides a complementary observational signature to the scalar-induced GWs discussed in Sec.~\ref{sec:PBHandGW}.
The measurement of spectral distortions could thus provide further constraints on the model parameters, particularly the non-minimal coupling $\xi$ and the coefficient of the $R^3$ term, $q$.

\section{Conclusion}
\label{sec:conc}
In this work, we have provided a systematic study of the Starobinsky-Higgs inflation model with an additional cubic term of the Ricci scalar, $R^3$. A key motivation for including this higher-order curvature term was to address the tension in the spectral index $n_s$ present in the standard Starobinsky-Higgs model when it is used to explain the dark matter abundance through PBH formation. Our analysis demonstrates that a negative $R^3$ term with $q \simeq -{\cal O}(10^{-5}$--$10^{-4})$ can successfully shift the spectral index from $n_s \simeq 0.94$ to $n_s \simeq 0.965$, bringing it well within the observationally-favoured region while maintaining the model's rich phenomenology.

Through both analytical and numerical treatments, we have shown that the $R^3$ term most strongly affects the first stage of the two-step inflationary scenario, primarily impacting the CMB-scale observations. For a negative coefficient $q$, the $R^3$ term flattens the curvature power spectrum near the CMB scale, raising the value of $n_s$. Simultaneously, the enhancement of curvature perturbations at smaller scales, crucial for PBH formation, remains largely unaffected by the presence of the $R^3$ term. This feature allows our model to produce PBHs in the mass range $10^{-16}M_\odot \lesssim M_{\rm PBH}\lesssim 10^{-11}M_\odot $ that could constitute the entirety of dark matter while being consistent with the latest CMB observations.

We have investigated various aspects of the model through concrete benchmark cases. The presented four benchmark cases which demonstrate how the model can achieve compatibility with CMB observations while producing distinct observational signatures. In particular, we find that for $\xi < 3/16$, the model predicts broader enhancement in the curvature power spectrum, leading to potentially observable CMB spectral distortions, while cases with $\xi = 5/16$ feature a $k^3$ growth, resulting in narrower enhancement.

Scalar-induced GWs have also been discussed for the benchmark cases with distinct characteristics determined by the parameters controlling the two phases of inflation. Cases 1a and 1b produce GW signals in the nano-Hertz frequency range, where recent observations by NANOGrav and other pulsar timing arrays have reported evidence for a stochastic GW background. Meanwhile, Cases 2a and 2b predict signals in the deci-Hertz range, accessible to future GW detectors like LISA, DECIGO, and BBO, offering additional opportunities for testing the model through GW physics.

The success of the $R^3$-corrected Starobinsky-Higgs model in reconciling CMB observations with PBH dark matter suggests that higher-order curvature terms could play an important role in early universe physics. Our analysis shows specific constraints on the model parameters: the $R^3$ coefficient is bounded as $-4\times 10^{-5} \lesssim q \lesssim -2\times 10^{-5}$ for the non-minimal coupling values $\xi \simeq 0.077$--$0.125$, while $q \simeq -1\times 10^{-4}$ is required for $\xi = 5/16$, with these ranges determined by both CMB compatibility and successful PBH formation.

Future observations will be crucial in testing these theoretical possibilities and constraining the model parameters. In particular, improved CMB measurements and the combination of PBH searches and GW observations could map the viable parameter space, including the coefficients of the higher-order curvature terms. The potential detection of spectral distortions by next-generation experiments like PIXIE would provide additional consistency checks of the model's predictions across different scales, offering a comprehensive probe of the early universe physics encoded in higher-order curvature terms.


\acknowledgments
We would like to thank Misao Sasaki and Seong Chan Park for useful discussions.
J.K. and Y.Z. were supported by National Natural Science Foundation of China (NSFC) under Grant No. 12475060 and by Shanghai Pujiang Program 24PJA134.
X.W. gratefully acknowledges the hospitality and support of the Kavli Institute of Physics and Mathematics of the Universe (Kavli IPMU), the University of Tokyo during her visit when the work was done. 
Y.Z. was also supported by the Fundamental Research Funds for the Central Universities, the Project 12047503 supported by NSFC, and Project 24ZR1472400 sponsored by Natural Science Foundation of Shanghai.
Z.R. was supported by NSFC under Grant No. 12035011.
Kavli IPMU is supported by World Premier International Research Center Initiative (WPI), MEXT, Japan.
This work was also supported in part by JSPS KAKENHI Grant No. JP24K00624, by Forefront Physics and Mathematics Program to Drive Transformation (FoPM), a World-leading Innovative Graduate Study (WINGS) Program, the University of Tokyo.


\bibliographystyle{JHEP}
\bibliography{refs}
\end{document}